\documentclass[11pt,twocolumn]{revtex4}
\usepackage{graphicx}

\newcommand{\be}{\begin{equation}}
\newcommand{\bea}{\begin{eqnarray}}
\newcommand{\ee}{\end{equation}}
\newcommand{\eea}{\end{eqnarray}}

\begin{document}
%\draft
%\begin{center}

\title{Dynamical stabilization of matter-wave solitons revisited}

\author{Alexander Itin$^{1,2,}$}

\author{Shinichi Watanabe$^1$, Toru Morishita$^1$}
%, Toru Morishita$^1$}
 \affiliation{ $^1$The University of
Electro-Communications, 1-5-1 Chofu-ga-oka, Chofu-shi, Tokyo 182-8585,
Japan and $^2$Space Research Institute, RAS, Profsoyuznaya Str. 84/32,
117997 Moscow, Russia }
% \date{\today}

\begin{abstract}

We consider dynamical stabilization of Bose-Einstein condensates (BEC)
by time-dependent modulation of the scattering length. The problem has
been studied before by several methods:  Gaussian variational
approximation, the method of moments, method of modulated Townes
soliton, and the direct averaging of the Gross-Pitaevskii (GP)
equation. We summarize these methods and find that the numerically
obtained stabilized solution has different configuration than that
assumed by the theoretical methods (in particular a phase of the
wavefunction is not quadratic with $r$). We show that there is
presently no clear evidence for stabilization in a strict sense,
because in the numerical experiments only metastable (slowly decaying)
solutions have been obtained. In other words, neither numerical nor
mathematical evidence for a new kind of soliton solutions  have been
revealed so far. The existence of the metastable solutions is
nevertheless an interesting and complicated phenomenon on its own. We
try some non-Gaussian variational trial functions to obtain better
predictions for the critical nonlinearity $g_{cr}$ for
metastabilization but other dynamical properties of the solutions
remain difficult to predict.
\end{abstract}

\maketitle

\section{Introduction}

The nonlinear Schrodinger equation (NLSE) appears in many models of
mathematical physics and has numerous applications. The one-dimensional
NLSE is famous due to its integrability and soliton solutions. The
two-dimensional and three-dimensional versions do not have such
properties and are much less explored.

In the last decade dynamics of BECs has attracted enormous amount of
interest which in turn is causing a renewed growth of interest in the
NLSE, since it is known that NLSE (often called the Gross-Pitaevskii
(GP) equation in that context) describes the dynamics of BEC at zero
temperature very well \cite{RefB1}.

While early analytical studies of BECs were concentrated on
(quasi-)one-dimensional systems, (quasi-)2D and 3D systems are more
important for real experiments. In 2D and 3D systems analytical
treatment of NLSE is very difficult and one has to use approximate
methods.

One of the very interesting and complicated phenomena being studied
recently is stabilization of BEC by the oscillating scattering length
in two and three dimensions.

In 1D geometry, bright solitons can be stable without trapping
potential if nonlinearity is attractive and sufficiently strong. In
NLSE with attractive interaction (corresponding to BEC with negative
scattering length) in 2D free space, kinetic energy can balance
interaction energy at certain critical value of nonlinearity $g_{cr}$,
but the resulting solution (Townes soliton) is unstable. That is, if
nonlinearity is either increased or decreased (and kept fixed
afterwards), the solution either expands or collapses correspondingly.
It was shown by several authors that stabilized solutions are possible
with the oscillating scattering length. The oscillations of the
scattering length lead to creation of pulsating condensate, i.e. some
kind of breather solution. One can draw an analogy with  Kapitza
pendulum (a pendulum with a rapidly oscillating pivot), where unstable
equilibria of unperturbed system is stabilized by means of fast
modulation. This idea was already applied to stabilization of beams in
nonlinear media \cite{Towers}. Among many other applications in related
fields, the atom wire trap suggested in Ref. \cite{Hau} should be
mentioned.
 In Refs. \cite{SU, AB} the novel application of
this stabilization mechanism to BEC physics was presented which in turn
encouraged several other works on that subject \cite{Abdullaev1,
Abdullaev2, Adhikari, Garcia,vektor}.

We consider here the problem of stabilization of BEC in 2D free space
by means of rapid oscillations of the scattering length in a greater
detail (the third dimension is assumed to be excluded from the
dynamics, say, due to a tight confinement). The system is described by
the GP equation: \be i \frac{\partial \psi}{\partial t} = - \frac{1}{2}
\nabla^2 \psi + \frac{\omega_r^2(t)}{2} r^2 \psi + g(t) |\psi|^2 \psi,
\label{GP} \ee where $ r^2=x^2+y^2 $ and $ g(t)=(8 \pi m \omega_z/\hbar
)^{1/2} N a(t)$ describes the strength of the two-body interaction. The
interaction $g(t)$ is rapidly oscillating: $g(t)=g_0+ g_1\sin(\Omega
t)$, while the confinement trap described by $\omega_r(t)$ is slowly
turned off. Refs. \cite{SU, AB, Abdullaev1, Abdullaev2} suggest it is
possible to obtain a dynamically stabilized bright soliton in free
space in such a way. Interactions between such objects were very
recently studied in Ref. \cite{vektor}. This is a very interesting
phenomenon not only in the context of BECs but also from a broader
scope of nonlinear physics.

 Such kind of stabilization in 3D has also been reported
\cite{Adhikari}. The latter finding is, however, in some disagreement
with other investigations on this topic (for example, Ref.
\cite{Abdullaev1}). In Ref. \cite{3D} it was shown that the scattering
length modulation may indeed provide for the stabilization in 3D, but
only in combination with a quasi-1D periodic potential. So 3D geometry
might need additional careful examination. In the present paper we
concentrate on quasi-2D case only, where also not everything is clear
yet. Unlike conventional 1D solitons, higher-dimensional solitonic
objects may decay. Therefore, it is interesting to investigate the
following question: is there indeed a novel genuine breather solution
behind the phenomenon of stabilization? As we show in this paper, it
turns out that the phenomenon does not fit into simple models being
suggested earlier. For theoretical description of the process, several
methods were used by different groups of authors: variational
approximation based on the Gaussian anzatz \cite{Abdullaev1,SU}, direct
averaging of the GP equation \cite{Abdullaev1}, a method based on
modulated Townes soliton \cite{Abdullaev1}, and the method of moments
\cite{Garcia}. Surprisingly, we find all the methods are not very
satisfactory even for qualitative predictions. In brief, the direct
averaging of the GP equation has the disadvantage of ommiting terms
which are of the same order as those responsible for creation of the
effective potential, while three other methods, although very
different, all rely on the unwarranted assumption of parabolic
dependence of the phase of the stabilized wavefunction on $r$: arg
$\psi = \alpha(t) + \beta(t) r^2$. We find that the behavior of the
exact numerical wavefunction is, however, completely different (see
Fig. \ref{f1efgh}). The above-mentioned parabolic approximation (PA) of
the phase factor  is very popular because it is appealingly simple and
indeed often appears in solutions of the time-dependent GP equation
\cite{Kagan}. Usually it comes from self-similar time evolution of the
condensate density, for example in 3D the following dynamics of the
condensate density is possible $ \rho(x,y,z)= [\lambda_1(t)
)\lambda_2(t)\lambda_3(t)]^{-1} \rho(x/\lambda_1(t), y/\lambda_2(t), z/
\lambda_3(t))$ , where coefficients $\lambda_i$ are coupled by
nonlinear differential equations. It is the important finding of the
present paper that in our problem a stabilized wavefunction does not
have such parabolic phase factor and does not fit into self-similar
patterns implied by the above-mentioned methods. This qualitative
difference between the exact numerical solution and all theoretical
models considered so far was not mentioned earlier. Besides, we noticed
presence of steady outgoing flux of atoms in numerical stabilized
solutions. So, even numerically there is no 2D soliton so far, but some
slowly decaying object instead. Section 2 reviews the abovementioned
theoretical methods. In Section 3 we give some results obtained using
the variational approximation with non-gaussian trial functions,
including "supergaussian anzatz". It is shown that a better accuracy
can be obtained for predicting critical nonlinearity $g_{cr}$, but we
were not able to determine accurately such dynamical properties as the
frequency of slow oscillations. Additionally, we checked the
supergaussian anzatz for another problem: determination of critical
number of attractive BEC in a parabolic well, and found it to be much
more accurate than the usual gaussian anzatz. This example also
demonstrates that the stabilization mechanism  is essentially more
complicated than that assumed by the present (PA-based) methods,
because predictions of the supergaussian anzatz for dynamical
properties of the stabilized solution are much less accurate than in
static problems.

 In Section 4 numerical results are presented and compared
with predictions of the theoretical methods discussed in Sections 2 and
3. Configuration of stabilized solution is discussed and dynamics of
some integral quantities of the solution is investigated.

In Section 5 concluding remarks are given. We mention the relation
between the BEC stabilization problem and stabilization of optical
solitons in a layered medium with sign-alternating Kerr nonlinearity.

\section{Several approximate methods to study the problem: PA-based methods (Gaussian variational approximation,
 the modulated Townes soliton, the method of moments), and the direct averaging of the GP equation.}
\subsection{PA-based methods}

\subsubsection{Gaussian variational approximation}

 The variational approach based on the Gaussian approximation (GA)
 is one of the most often used in studying dynamics of the GP equation.
 In actual calculations  this approximation
however often gives a large error as compared to exact numerical
results \cite{Metens,Abdullaev2}. For example, in Ref. \cite{Metens}
the Gaussian approximation in dynamics of attractive BEC  was compared
to exact numerical solution of the GP equation. It was found that in
estimating the critical number ${\cal N}_c$ of the condensate (the
maximal number of condensed particles in a trap before collapse occurs)
the Gaussian approximation gives a 17\% error, and similar values of
discrepancy for other dynamical quantities (as a useful test, in the
Appendix we provide  corresponding results obtained with a
supergaussian variational ansatz). However, it seems that in this
example GA enables to reproduce important features of the system at
least qualitatively. The GA was also used in many other treatments of
the GP equation using a variational technique. In particular, it was
applied to the problem of BEC stabilization by the oscillating
scattering length. The Lagrangian density corresponding to the GP
equation (\ref{GP}) is

\be L[\psi]=\frac{i}{2} \left( \frac{\partial \psi}{\partial t} \psi^*
- \frac{\partial \psi^*}{\partial t } \psi \right) - \frac{1}{2} \left|
\frac{\partial \psi}{\partial r} \right|^2 - \frac{1}{2}g(t)|\psi|^4
\label{GPden}. \ee

The normalization condition for the wavefunction is $2\pi \int_0^\infty
|\psi|^2 r dr=1$.

In Ref. \cite{SU}, a variational method with the following Gaussian
anzatz was used, \be \psi(r,t) = \frac{1}{\sqrt{\pi} R(t)} \exp \left[
-\frac{r^2}{2R^2(t)} + i \frac{\dot R(t)}{2R(t)} r^2 \right],
\label{gaussian} \ee where $R(t)$ is the variational parameter that
characterizes the size of the condensate, and the phase factor of the
wavefunction describes the mass current \cite{SU,AB,Cirac}.

After substitution of expression (\ref{gaussian}) into the Lagrangian
density (\ref{GPden}) one obtains the effective Lagrangian $L = 2 \pi
\int_0^{\infty} r L[\psi] dr $ and the corresponding Euler-Lagrange
equations of motion. One can obtain then
 the equation of motion for $R(t)$ as \be
\ddot{R}(t) = \frac{1}{R^3(t)} + \frac{g_0+g_1 \sin{\Omega t}}{2 \pi
R^3 (t)}. \ee So the gist of the model is to represent the 2D BEC as a
classical nonlinear pendulum with modulated parameters. It is important
that other one-parameter PA-based anzatzes also give the same nonlinear
pendulum ($\ddot{R}=(a+b \sin \Omega t)/R^3$, where $a,b$ depend on the
parameters $g_1,g_0,\Omega$), but with different functional dependence
of a,b on the parameters.

 The authors of Ref. \cite{SU} use then the Kapitza averaging
method to study behavior of the system with the rapidly oscillating
scattering length. They assume the dynamics of $R$ can be separated
into a slow part $R_0$ and a small rapidly oscillating component
$\rho$: $R=R_0(t) + \rho(\Omega t) $. From the equations of motion for
$R_0$ and $\rho$ one extracts the effective potential for the slow
variable $U(R_0) \approx \frac{A_2}{R_0^2}+ \frac{A_6}{R_0^6} $ and
determines its minimum \be R_{min} = \left(\frac{-3}{4 \pi (g_0+2
\pi)}\right)^{1/4} \left(\frac{g_1}{\Omega}\right)^{1/2} \label{rmin}.
\ee

From the expression for the effective potential for $R_0$ they obtained
dependence of the monopole moment $<r>$ and the breathing-mode
frequency $\omega_{br}$ on parameters $g_1, \Omega$. The frequency of
small oscillations (breathing mode) around the minimum is given by
\cite{SU} \be \omega_{br}^2= \frac{8 \Omega^2}{3 g_1^2} (g_0+2 \pi)^2.
\label{ombr}\ee

Their numerical calculations were done for $g_0= -2 \pi$. One can see
that theoretical predictions (\ref{rmin}) and (\ref{ombr}) based on the
Gaussian approximation can catch $(g_1/\Omega)^{1/2}$ dependence of the
monopole moment $<r>$ and $(\Omega/g_1)$ dependence of the
breathing-mode frequency $\omega_{br}$ but cannot determine the
corresponding coefficients of proportionality, of which the one in
(\ref{rmin}) becomes infinity while the one in (\ref{ombr}) becomes
zero for $g_0=-2 \pi$, the value actually used in the numerical
calculations. On the other hand, from numerical calculations they were
able to determine the coefficients as $1.06$ and $0.32$ correspondingly
(see Fig. 2 of Ref. \cite{SU}). It was also determined in Ref.
\cite{SU} that in order to stabilize the bright soliton, $|g_0|$ must
exceed the critical value of collapse $|g_{cr}|$. Their numerical
estimate for $|g_{cr}|$ is $\approx$ 5.8 while theoretical estimate
based on Gaussian approximation is $2\pi \approx 6.28$. The $2 \pi$
estimate in fact corresponds to fitting the so-called Townes soliton by
a Gaussian trial function as will be discussed below.

Inspired by the idea of comparing a numerical solution with simple
model nonlinear pendulum, one may ask if it is possible to obtain a
better accord with the numerical experiments using different ansatzes.
We study this question in Section 3, and it seems that only the
stationary Townes soliton can be fit accurately, but not the stabilized
breather solutions.

\subsubsection{Modulated Townes soliton}

A method based on modulated Townes soliton used in  Ref.
\cite{Abdullaev1,Garcia} should be mentioned. The Townes soliton is a
stationary solution to the 2D NLS equation with constant nonlinearity
$g_{cr}$. In our notations $|g_{cr}| \approx 1.862 \pi \approx 5.85$.
This solution is unstable: if $|g|$ is slightly increased or decreased,
the solution will start to collapse or expand correspondingly. If the
value of $g$ is close to $g_{cr}$, one may search for a solution of the
problem with  fast oscillating $g$ in the form of a modulated Townes
soliton, as described in Refs. \cite{Abdullaev1,Garcia}. A solution is
sought in the form of \bea \Psi(r,t) &\approx& [a(t)]^{-1} R_T[r/a(t)]
e^{iS}, \nonumber\\  \quad S &=& \sigma(t) + \frac{r^2 \dot{a} }{4a},
\quad \dot{\sigma}=a^{-2}, \label{mT} \eea where $R_T$ represents
amplitude of the Townes soliton. Then, starting from the approximation
(\ref{mT}), one can derive the evolution equation for $a(t)$ and so
determine the dynamics of the system. Note that the approach is also
PA-based. It is inevitable if we are to use one-parameter self-similar
trial function in the  form of $|\psi(r,t)|=A f(r/a,t)$.

\subsubsection{The method of moments}
Another PA-based method we would like to mention here is the method of
moments \cite{Garcia}. One introduces integral quantities
$I_1,I_2,I_3,..$ as \bea I_1 &=&
\int_0^{\infty} |\psi|^2 d {\bf r}, \qquad I_2 = \int_0^{\infty} r^2 |\psi|^2 d {\bf r},  \nonumber\\
I_3 &=& i \int_0^{\infty} \left(\psi \frac{\partial \psi*}{\partial r} - \psi^* \frac{\partial \psi}{\partial r}  \right) r d {\bf r}, \\
I_4 &=&  \frac{1}{2}\int_0^{\infty} \left( |\nabla \psi|^2 + \frac{n}{2} g(t) |\psi|^4 \right) d {\bf r}, \nonumber\\
I_5 &=& \frac{n}{4} \int_0^{\infty} |\psi|^4 d {\bf r}, \nonumber
 \eea
where $n=2,3$ is the dimension of the problem. In 2D, $d {\bf r} = 2
\pi rdr$, and in 3D $d {\bf r} = 4 \pi r^2 dr$.

For all t, we have $I_1= 1$. For the remaining $I_i$  one can write
down the dynamical equations of motion as \cite{Garcia}:

\bea \dot{I_2}=I_3, \quad \dot{I_3}=I_4, \quad \dot{I_4}= g
\frac{n-2}{n} I_5 + \dot{g} I_5, \label{moments}\\
\dot{I_5}= \frac{n \pi 2^n}{8} \int_0^{\infty} \frac{ \partial
|\psi|^4}{\partial r} \frac{\partial \arg \psi}{\partial r} r^{n-1} dr
 \eea
The system of equations for the momenta is not closed because of $I_5$,
and one should make some approximation in order to close it. In Ref.
\cite{Garcia} it was assumed that \be \arg \psi = \frac{I_3 r^2}{4
I_2}, \label{argu} \ee i.e. the phase factor is proportional to $r^2$
(so that again it is a PA-based method) and the coefficient of
proportionality is given by the ratio of $I_3$ and $I_2$. Then the
system (\ref{moments}) poses dynamical invariants \cite{Garcia}:
\bea Q_1 &=& 2(I_4 - gI_5)I_2 - \frac{1}{4} I^2_3 , \\
Q_2 &=& 2I^{ n/2}_2 I_5.  \label{quinv}
 \eea
With the help of these invariants, the system becomes \be \ddot{I}_2 -
\frac{1}{2I_2} \Bigl( \dot{I}_2 \Bigr)^2 = 2 \left( \frac{Q_1}{I_2} + g
\frac{Q_2}{I^{n/2}_2} \right) . \ee Introducing $X(t)=\sqrt{I_2(t)}$
one obtains \cite{Garcia} \be \ddot{X}= \frac{Q_1}{X^3} + g(t)
\frac{Q_2}{X^{n+1}}.  \label{xmodel} \ee

The equation is analogous to that obtained by other PA-based methods.
One can investigate the obtained equation (\ref{xmodel}) using various
methods of nonlinear dynamics. The simplest Kapitza averaging method
can be used again, but of course it is better to use rigorous averaging
technique since modern averaging methods are available \cite{AKN} which
have been extensively used already in plasma physics, hydrodynamics,
classical mechanics \cite{It1}. The authors of Ref. \cite{Garcia}
fulfilled rigorous analysis of model Eq. \ref{xmodel} using results of
Ref. \cite{Ref_Garcia}. It is important to have in mind that the
relation between the exact dynamics of the full system and that of the
model (\ref{xmodel}) of the method of moments remains unclear,
therefore one cannot determine sufficient conditions for stabilization,
etc. In Ref. \cite{Garcia} it was noticed that the correspondence
between numerical simulation of full 2D GP equation and dynamics of the
model system (\ref{xmodel}) is not good. As it is seen from Fig. 3 of
Ref. \cite{Garcia}, neither the frequency of slow oscillations nor the
position of the minimum of the effective potential is predicted
correctly. Nevertheless, we found that in numerical stabilized
solutions magnitudes of $Q_1$ and $Q_2$ are often well-conserved, i.e.
they oscillate about some mean value (see Section 4).

\subsection{Direct averaging of the GP equation}
 Ref. \cite{Abdullaev1} also explores the Gaussian variational
approximation. Beside that, a very promising method of directly
averaging  the GP equation was investigated. It is based on an
analogous method used for the one-dimensional NLSE with periodically
managed dispersion
 (in the context of optical solitons) \cite{Kath}. In Ref.
\cite{Abdullaev1} the solution is sought as an expansion in powers of
$1/\Omega $ (in our notation): \be \psi(r,t) = A (r,T_k) + \Omega^{-1}
u_1(A,\zeta) + \Omega^{-2} u_2(A,\zeta) + ....,  \label{expansion} \ee
 with $<u_k>=0$, where $< ... >$ stands for the
average over the period of the rapid modulation, $T_k \equiv
\Omega^{-k} t $ are the slow temporal variables ($k=0,1,2, . . .$),
while the fast time is $ \zeta = \Omega t$. Then, for the first and
second corrections the following formulas were obtained: \bea u_1 &=& -
i [\mu_1 - <\mu_1>]| A|^2 A, \nonumber\\    \quad \mu_1
& \equiv & \int_0^{\zeta}[g(\tau)-<g_1>] d \tau, \\
u_2 &=& [\mu_2 - <\mu_2>][ 2i |A|^2  A_t + i A^2 A^{*}_t +
\Delta(|A|^2 A)] \nonumber\\ &-& |A|^4 A ( \frac{1}{2}[(\mu_1-<\mu_1>)^2-2M] + \nonumber \\
&+&<g>(\mu_2 - <\mu_2>)) ],  \nonumber\\
\quad \mu_2 &=& \int_0^{\zeta} (\mu_1-<\mu_1>) ds, \quad M  =
\frac{1}{2} (<\mu_1^2> - <\mu_1>^2). \nonumber\eea Using these results,
the following equation was obtained for the slowly varying field
$A(r,T_0)$, derived up to the order of $\Omega^{-2}$: \bea -i
\frac{\partial A}{\partial t} &=& \Delta A + |A|^2 A + 2 M \left(
\frac{g_1}{\Omega} \right)^2 [|A|^6 A   \nonumber \\
 - 3 |A|^4 \Delta A &+&
2 |A|^2 \Delta (|A|^2 A) + A^2 \Delta (|A|^2 A^*) ]. \nonumber \\
\label{slow} \eea The above equation was represented in the
quasi-Hamiltonian form \bea \Bigl[ 1 &+& 6 M \left(
\frac{g_1}{\Omega}^2  |A|^4\right) \Bigr] \frac{\partial A}{\partial t
} = - \frac{\delta H_q}{\delta A^*}, \nonumber\\  H_q &=& \int dV
\Bigl[ |\nabla A|^2 - 2 M \left(
\frac{g_1}{\Omega} \right)^2 |A|^8  \nonumber\\
 &-& \frac{1}{2} |A|^4 + 4M \frac{g_1}{\Omega} |\nabla (|A|^2 A )|^2 )^2\Bigr].
 \label{quasiham}
\eea

 However, some contribution was missed while deriving
 Eq. (\ref{slow}). Let us take into account the third correction
 $u_3(A,\zeta)$:
\bea \psi(r,t) &=& A (r,T_k) + \Omega^{-1} u_1(A,\zeta) + \Omega^{-2}
u_2(A,\zeta)   \nonumber\\ &+& \Omega^{-3} u_3(A,\zeta)+ .... \eea
 Then, up to terms of order  $\Omega^{-2}$
 it changes nothing in r.h.s of Eq.(\ref{slow}) (spatial part), but it
 adds to l.h.s. of  Eq. (\ref{slow}) an undetermined term $\Omega^{-2} \partial u_3/ \partial
 \zeta$ . This term has the same order $\Omega^{-2}$ as the terms from
 the second correction.  So we do not get here a consistent equation for the slow field $A$ because
   we do not have a closed set of equations for the second-order corrections (third order correction becomes
  second order correction after differentiating in time), and so the
quasi-Hamiltonian (\ref{quasiham}) contains an undetermined
 error of the second order in $\Omega^{-1}$.
 The influence of the contribution is not very clear but require additional investigation.
 Nevertheless, formally the omitted terms have the same order as those responsible for the creation
of the effective potential. Having in mind how many difficulties arise
in averaging of systems of {\em ordinary} differential equations
\cite{AKN}, the rigorous direct averaging of the GP equation
constitutes a very interesting and challenging open problem, since in
principle it could reveal a true periodic solutions in such oscillating
objects.
%It seems that only
%this method is capable to determine  solitonic

\section{Variational approximation with non-Gaussian ansatzes}

Here we try to investigate the system more accurately using some
non-Gaussian ansatzes and see if it is possible to get more accurate
theoretical estimates. One may be interested in three dynamical
quantities of the system: the value of critical nonlinearity $g_{cr}$,
slow frequency of breathing oscillations of the stabilized soliton
$\omega_{br}$, and minimum of the effective potential $R_{min}$ about
which the expectation value of the monopole moment $<r>$ oscillates
slowly.

%It turns out that only $g_{cr}$ can be obtained accurately if one goes
%beyond the Gaussian approximation still remaining within the parabolic
%approximation.

 Table 1 summarizes results of variational predictions for the critical
nonlinearity $g_{cr}$ and frequency of small breathing oscillations
using several different ansatzes. Note that the phase dependence of a
one-parameter trial function is not important for calculating $g_{cr}$.
It is understood that if we choose a trial wavefunction with its
amplitude in the form of $|\psi(r,t)| = Af[r/a(t)]$, then we need to
use a phase factor with quadratic $r-$ dependence in order for the
ansatz to be self-consistent (i.e., the mass current generated by the
changing parameter would be incorporated in the phase factor of an
ansatz). On the other hand, since amplitude part of the trial function
is just an approximation, one may try to use other forms of phase
factor with the same functional form of the amplitude.

% However, in order to use non-quadratic phase factors
%self-consistently, one need either to use many-parameter not
%self-similar trial functions, or include a parameter in a different way
%as compared to usual functional form $f[r/a(t)]$.

When predicting the frequency of breathing oscillations from the
corresponding effective potential, it is easy to obtain the result for
small amplitude linear breathing oscillations (given in Table 1), but
in actual stabilized solutions amplitudes of breathing oscillations are
not so small.

 It is possible to take into account anharmonicity of breathing oscillations.
As was mentioned earlier, all PA-based anzatzes produce the nonlinear
pendulum $\ddot{R}+(a+b \sin \Omega t)/R^3$, with a corresponding
effective potential having $ R_{min} = \left(- \frac{3b^2}{2 \Omega^2
a} \right)^{1/4}$ , $ \omega_{br}= \sqrt{\frac{8}{3}} \Omega |a/b|,$
where $\omega_{br}$ is the frequency of the small amplitude breathing
oscillations (near the bottom of the effective potential). For larger
breathing oscillations the (anharmonic) breathing frequency will be
amplitude-dependent: $ \omega_{br}^{anh} = 2 \pi \left( \sqrt{-\frac{
2}{h}} \left[ \frac{x_3}{\sqrt{x_2-x_3}} \mbox{\bf K}(k) +
\frac{x_2}{x_1} \sqrt{x_2-x_3} \mbox{\bf E}(k) \right] \right)^{-1}$,
with $\quad k=\sqrt{\frac{x_2-x_1}{x_2-x_3}},$ where $x_1=R_1^2$,
$x_2=R_2^2$ ($R_1,R_2$ being the turning points), $x_3$ is the third
root of the equation $ h = \frac{a}{2x}+\frac{b}{4 \Omega^2 x^3}$. The
magnitudes of $x_1,x_2,x_3,h$ can be determined from numerically
obtained breathing oscillations (but results depend on the choice of a
particular anzatz).  Even this improvement is not helpful, simply
because the parabolic approximation is not valid.

Finding $g_{cr}$ only might be considered as an approximation to the
stationary Townes soliton  by a trial function so that the mass current
term equals zero and that a phase factor may be skipped from the
calculations. It is known that the Townes soliton $\psi_t=e^{it}
R_T(r,t)$ at large $r$ has asymptotic behavior for its amplitude in the
form $ R_T \sim e^{-r}/\sqrt{r} $ . So that Gaussian ansatz is not very
good for finding $g_{cr}$ just because it is decaying too fast at large
$r$. The supergaussian trial function  provide a better approximation,
namely $g_{cr}=\pi 2^{\frac{1}{\mbox{ln} 2}}  \mbox{ln}2 $ which
corresponds to the supergaussian wavefunction with $\eta=\eta_T= 2
\mbox{ln} 2 < 2$. Previously the supergaussian ansatz was used to fit
stationary solutions of some nonlinear problems including NLS equation
in the context of BECs \cite{Pramana}. The superposition of two
Gaussians in the form $ A \exp(- \frac{r^2}{2R^2}) \mbox{Cosh}(\gamma
\frac{r^2}{2R^2})$ also enables one to obtain some improvement: $g_{cr}
\approx 5.883$. The Secanth ansatz $$ \psi=  \frac{A}{\cosh(r/R)} \exp
[i S(\dot{R},R) r^2 ] $$ works better, with only one parameter it
overcomes the above-mentioned two-parameter trial functions. A very
good approximation is provided by the simplest ansatz among all
considered: \be \psi=  \frac{1}{3 R \sqrt{\pi}} \left(1+\frac{r}{2R}
\right ) \exp \left\{-\frac{r}{2R} + i S(\dot{R},R)r^2 \right\}.
\label{exponent}\ee
 It fits the Townes soliton adequately both at the
origin and asymptotically at infinite $r$ ( a pre-exponential
multiplier is not so important as the exponential factor ).
 The pre-exponential factor is needed in order to fulfill the boundary
 condition in the origin $ \lim_{r \to 0}\frac{1}{r}\psi_r < \infty$
.  Note that in the supergaussian ansatz the former condition is not
fulfilled, otherwise (if one included it in a similar way) the result
would be better at the cost of more bulky calculations. The accuracy of
the prediction implies that ansatz (\ref{exponent})  provides a very
good approximation to the Townes soliton at fixed $R$, and could
approximately represent the modulated Townes soliton when  $R$ is
time-dependent and the phase factor with parabolic $r-$dependence is
used in accordance with the continuity condition.

\begin{table*}[]
\caption[]{Variational predictions for the properties of stabilized
solutions.}
\bigskip
\begin{tabular}{|@{\quad} c @{\quad}|@{\quad} c@{\quad}| @{\quad}c@{\quad}| @{\quad}c@{\quad}| @{\quad}c@{\quad}| }
\hline Ansatz   &  Amplitude part    & $g_{cr}$, & $g_{cr}$, & $\kappa_{br}$    \\
  & of the anzatz   &   analytical  & approximate &(linear prediction,\\
 &               &  expression  &  value  & $\omega_{br}=\kappa_{br} \Omega/g_1$)\\

\hline
 Gaussian  & $A \exp(-\frac{r^2}{2R^2})$ & 2 $\pi$   &6.283  & $\sqrt{\frac{8}{3}} (g_0+2\pi)$ \\
 Supergaussian & $ A \exp(-\frac{1}{2} \Large(\frac{r}{R} \Large)^{\eta})$  &  $\pi  2^{\frac{1}{\mbox{ln} 2}} $ \mbox{ln}2  &
 5.919 &  \\
 Secanth  & $ A \mbox{Sech} \Large( \frac{r}{R} \Large)$ & $ 2\pi \ln{2} \frac{2\mbox{ln}2+1}{4 \mbox{ln}
 2-1}$  & 5.863 & $\sqrt{\frac{8}{3}}\frac{| 2 \ln 2 +1 + g_0 (\frac{4 \ln 2 -1}{2 \pi \ln 2}) | }{\frac{4\ln2-1}{2 \pi \ln 2} }$     \\
 Exponential & $A (1+\frac{r}{2R}) \exp(-\frac{r}{2R})$ & $\frac{144}{77}  \pi$ &
 5.875 & \\
\hline
\end{tabular}
\end{table*}

%\begin{table*}[]
%\caption[]{Variational predictions for the value of $g_{cr}$}
%\bigskip
%\begin{tabular}{|@{\quad} c @{\quad}|@{\quad} c@{\quad}| @{\quad}c@{\quad}| @{\quad}c@{\quad}|}
%\hline Ansatz   &  Amplitude part    &  Analytical expression for
%$g_{cr}$ &
%Approximate value   \\
%\hline
% Gaussian  & $A \exp(-\frac{r^2}{2R^2})$ & 2 $\pi$   &6.283
% \\ Supergaussian & $ A \exp(-\frac{1}{2} \Large(\frac{r}{R} \Large)^{\eta})$  &  $\pi  2^{\frac{1}{\mbox{ln} 2}} $ \mbox{ln}2  &
% 5.919
% \\  Double-gaussian & $ A \exp(- \frac{r^2}{2R^2}) \mbox{Cosh}(\gamma \frac{r^2}{2R^2})$ & complicated  &
% 5.883
% \\  Secanth  & $A \mbox{Sech} \Large( \frac{r}{R} \Large)$ &$ 2\pi\frac{\mbox{ln}2+1}{4 \mbox{ln}
% 2-1}$
% &5.865
% \\  Exponential & $A (1+\frac{r}{2R}) \exp(-\frac{r}{2R})$ & $\frac{144}{77}  \pi$ & 5.875
% \\
%\hline
%\end{tabular}
%\end{table*}

After obtaining estimates for $g_{cr}$, one can use the above-mentioned
ansatzes in order to find an effective potential, its minimum and
frequency of the breathing oscillations of the monopole moment about
this minimum in the same way as it was done for the Gaussian ansatz. We
checked the Sech ansatz and the supergaussian with quadratic phase
dependence. In the supergaussian ansatz  the parameter $\eta$ was fixed
at the value of its "Townes soliton-like" solution $\eta=\eta_T=2
\mbox{ln} 2$. In such a way the variational approximation with
supergaussian ansatz resembles method of modulated Townes soliton.
However, we find that such trial function seriously underestimate
minimum of the effective potential (i.e. the mean value about which the
monopole moment oscillates). Nevertheless, the result of the Gaussian
ansatz is even worse since for $g_0=2 \pi$ it gives the diverging
expression for $R_{min}$ and zero for frequency of slow breathing
oscillations $\omega_{br}$, as mentioned in Section 1 and \cite{SU}. A
natural idea for remedy is to use two-parameter trial functions to
reproduce the non-parabolic phase factor dependence on $r$. In the
supergaussian ansatz it can be done by considering $\eta$ as a
dynamical (time-dependent) parameter. The problem is that it is
difficult to obtain the self-consistent expression for the phase
factor. We also try the supergaussian ansatz with fixed $\eta$ and with
non-quadratic phase dependence (which is unfortunately not
self-consistent trial function) $ \psi(r,t) = A \exp \left[
-\frac{(a+ib) r^{\eta_T}}{2} \right], \label{imgaussian} $ where
$A,a,b,$ and $\eta$ are all functions of time, parameter $\eta$ is
fixed at the value of its Townes soliton-like solution $\eta=\eta_T=2
\mbox{ln} 2$. We find that such modification drastically changes
dynamical parameters of the system. Still, the resulting model is the
same classical nonlinear pendulum as in the Gaussian approximation, but
with different parameters. The rigorous way to employ the two-parameter
supergaussian ansatz is to let $\eta$ be a dynamical variable and
construct a phase factor fulfilling continuity condition for the trial
function. One could then obtain the two-dimensional effective potential
within the same Kapitza approach.

As a useful test of applicability of the supergaussian anzatz, we
determine the critical number of attractive BEC in the 3D parabolic
trap studied in Ref. \cite{Metens}. Their numerical result was
$N_{cr}=1258.5$, while the gaussian approximation yields
$N_{cr}^{G}=1467.7$. We found the supergaussian prediction to be very
accurate $N_{cr}^{SG}=1236.1$.

\section{Numerical results}

Numerical calculations reveal the fact that stabilized solutions  do
not have parabolic phase factors in contradiction to all the methods
considered in Section 2 (except the method of direct averaging). The
calculations were done using explicit finite difference schemes. We use
explicit finite differences of second and forth order for spatial
derivatives and 4-th order Runge-Kutta method for time propagation. We
use meshes varying from 2000 to 10000 points, timesteps $\Delta t =
0.0001 \sim 0.0004 $, and spatial steps $\Delta r = 0.02 \sim 0.04$. In
addition, we found that it is very important to use absorbing
(imaginary) potential at the edge of the mesh, in accordance with the
conclusions of Ref. \cite{Garcia}. Without such an adsorbing potential,
a wave reflected from the edge sometimes destroys the otherwise stable
solution.
% The absorbing potential was included in the
%form of an imaginary potential $V_a=i V_0 \theta()$

\begin{figure*}
% \centering
{\includegraphics[width=6cm]{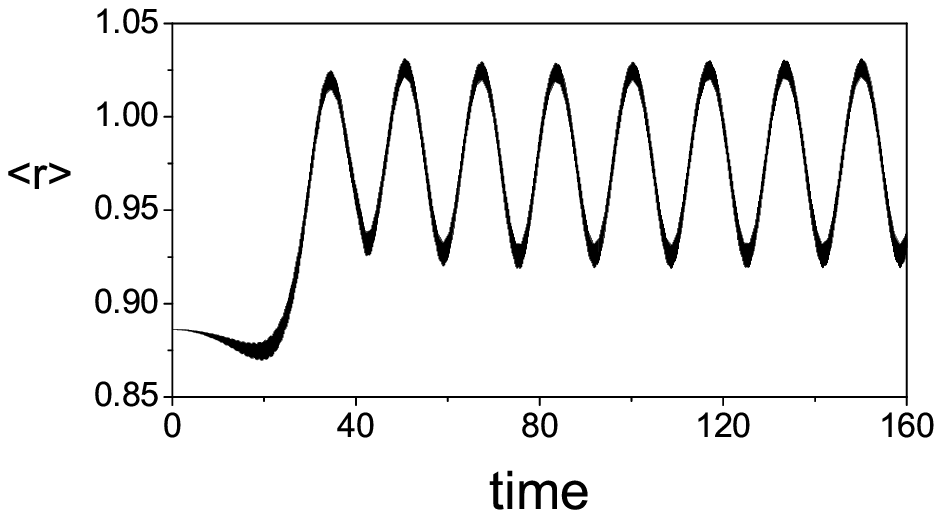} (a) }
 { \includegraphics[width=6cm]{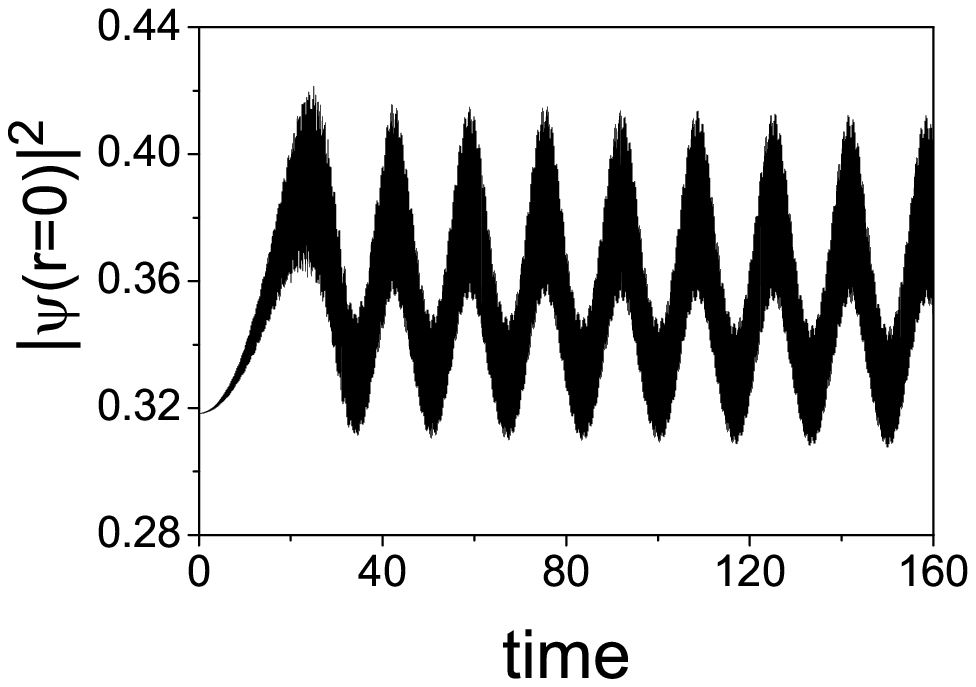} (b) }
 { \includegraphics[width=6cm]{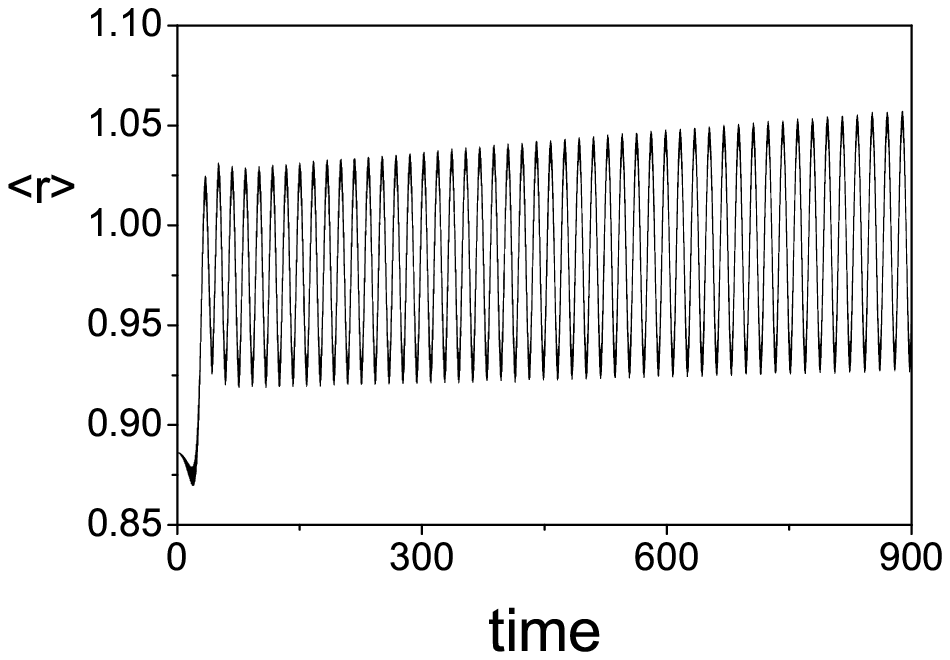} (c)  }
{ \includegraphics[width=6cm]{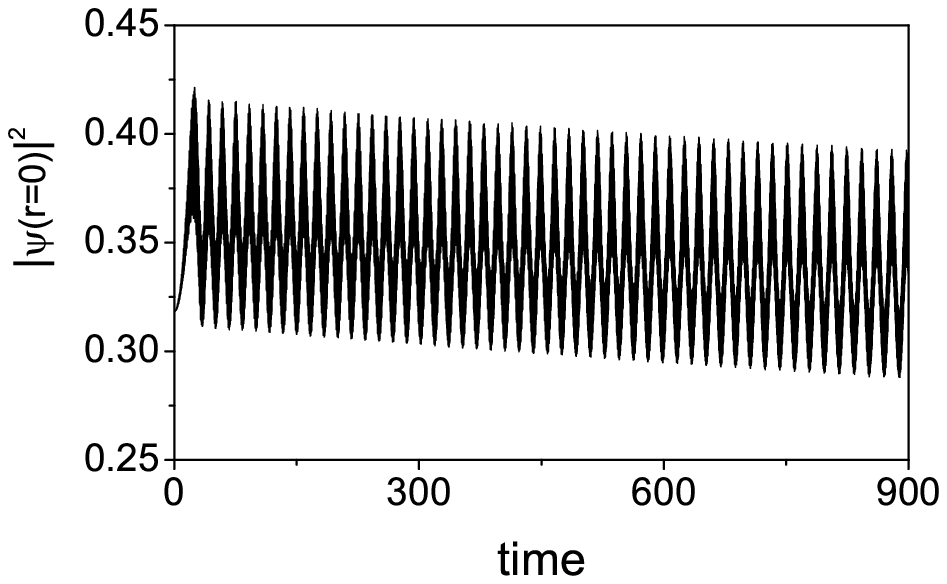} (d) }
{\includegraphics[width=6cm]{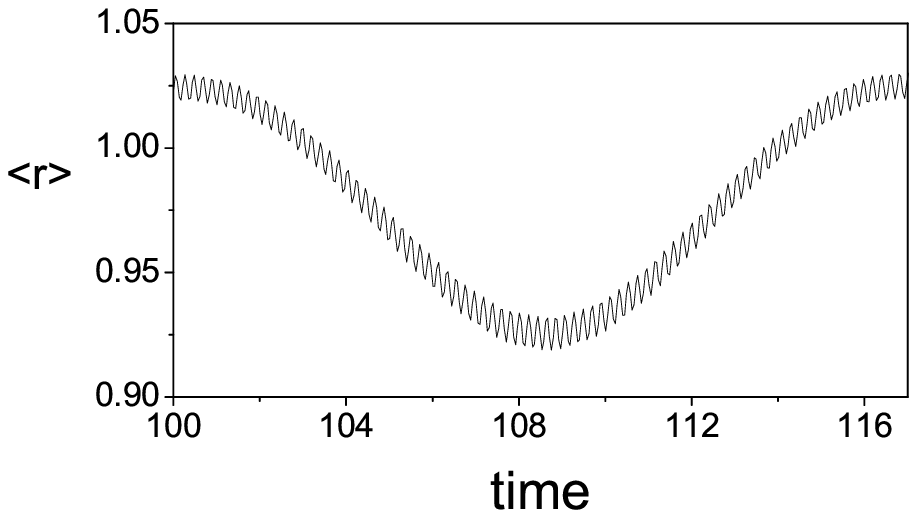}(e)}
 \caption{(a)Oscillations of the monopole moment after turning off the
 trap. Parameters are  $g_0=-2 \pi$, $g_1=8 \pi$, $\Omega=30$. The trap was turned off completely
 at $T_{off}=30$. (b) Time evolution of amplitude of wavefunction at the origin.
 (c) Oscillations of the monopole moment on longer time scale. (d) Time evolution of amplitude of wavefunction at the origin on
 longer times. (e)  The oscillations of the monopole moment from previous figure on finer scale.
 Tiny high frequency oscillations are seen. }
\label{f1}
\end{figure*}

 Following \cite{SU}, initially we start with a
Gaussian wavepacket in a parabolic trap. Then the trap was slowly
turned off while the oscillating nonlinearity was slowly turned on in a
 way similar to Ref. \cite{SU}.
 In Figure \ref{f1} one can see indeed the creation
of a stabilized soliton. In Figure \ref{f1}b and \ref{f1}d oscillations
of amplitude of the wavefunction at the origin are shown. It decays
very slowly. In fact, this is in accord with the calculations of Ref.
\cite{SU}: after a careful examination of the corresponding figures in
that paper one notices the same behavior. Monopole moment grows very
slowly (Figs. \ref{f1}a,c). We checked that in the case when the trap
is not turned off completely, the norm is conserved during the same
long time with a high accuracy (of order $10^{-8}$), so decay is
certainly not due to numerical errors.

\begin{figure*}
 \centering
{ \includegraphics[width=6cm]{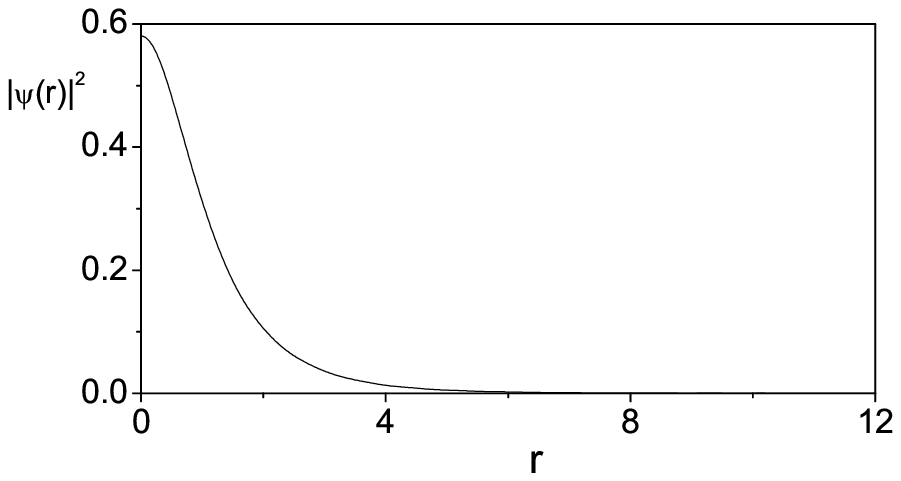} (a)}
 { \includegraphics[width=6cm]{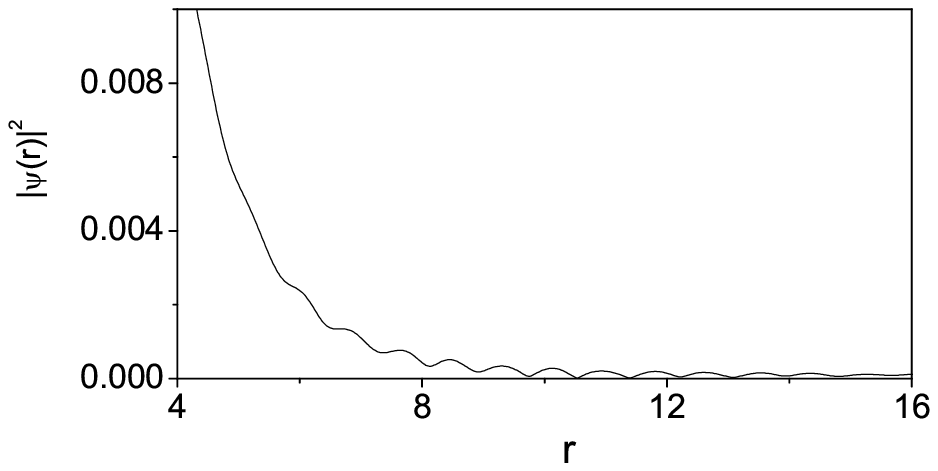} (b)}
  { \includegraphics[width=6cm]{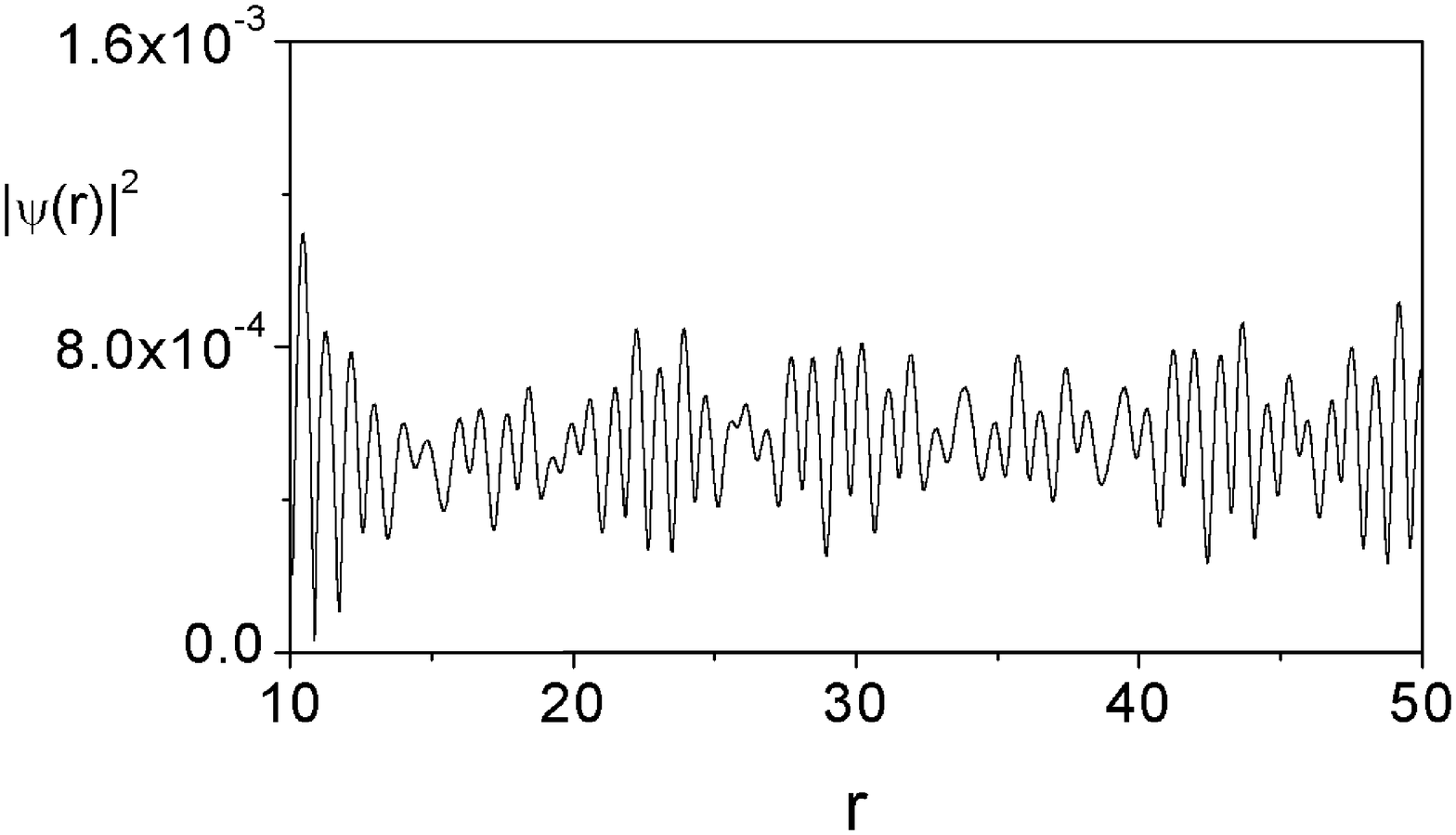} (c)}
{ \includegraphics[width=6cm]{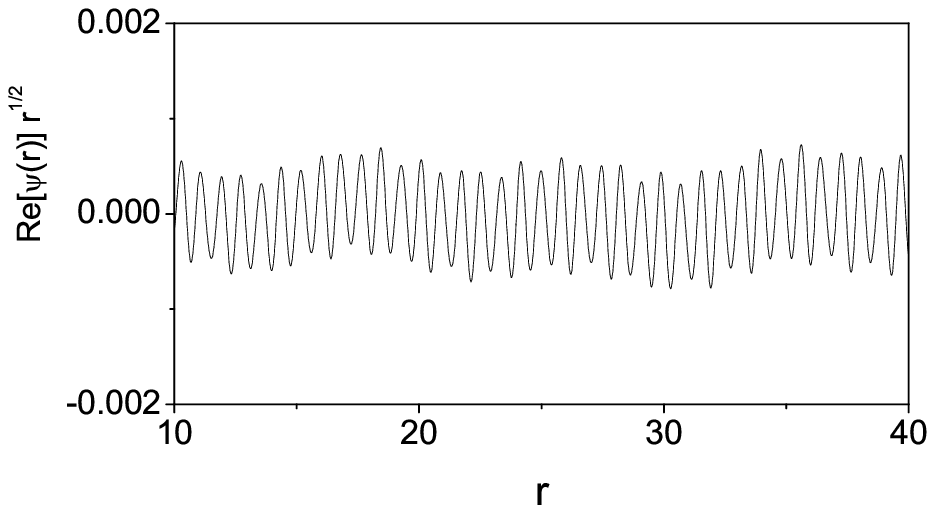} (d)}
{\includegraphics[width=6cm]{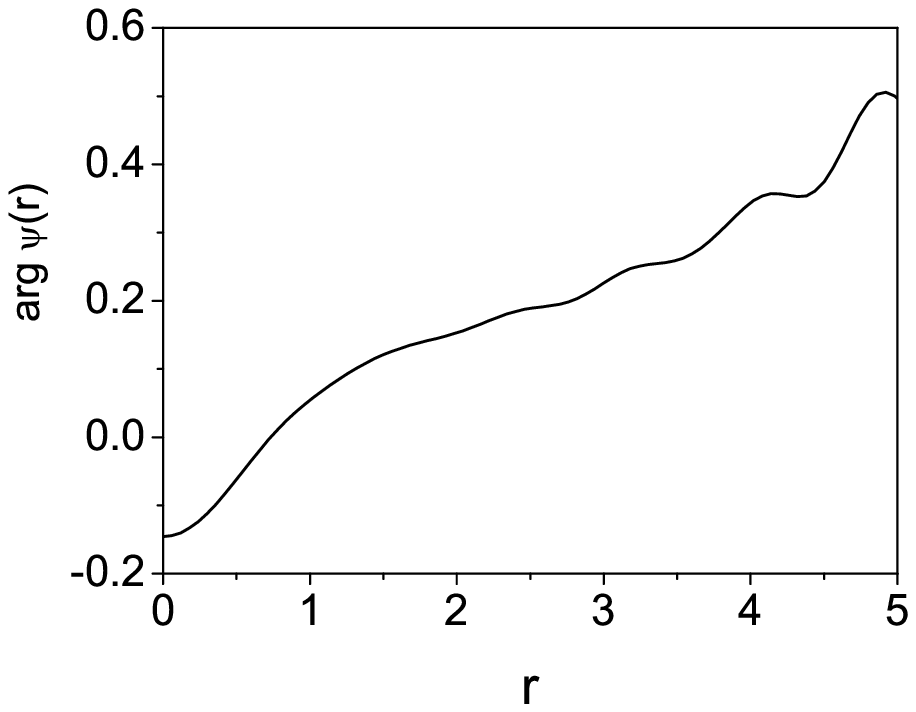} (e)}
 {\includegraphics[width=6cm]{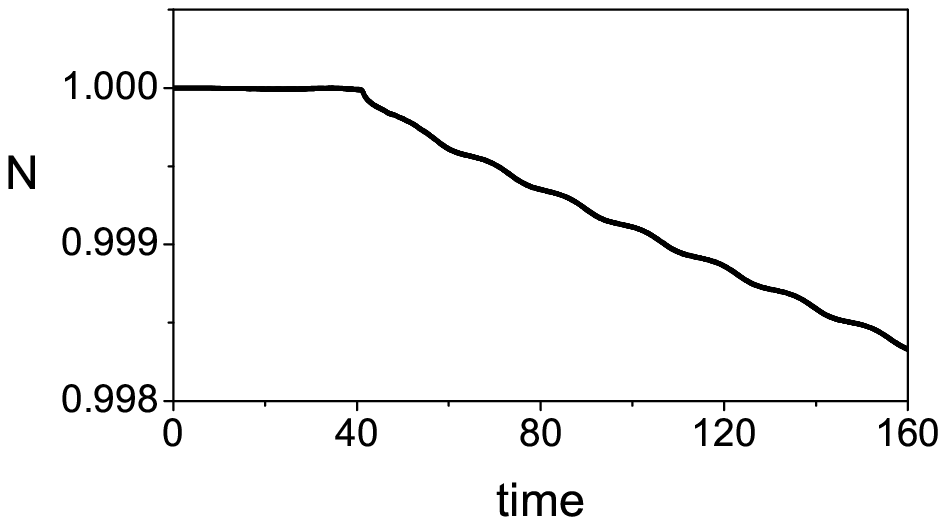} (f)}
{\includegraphics[width=6cm]{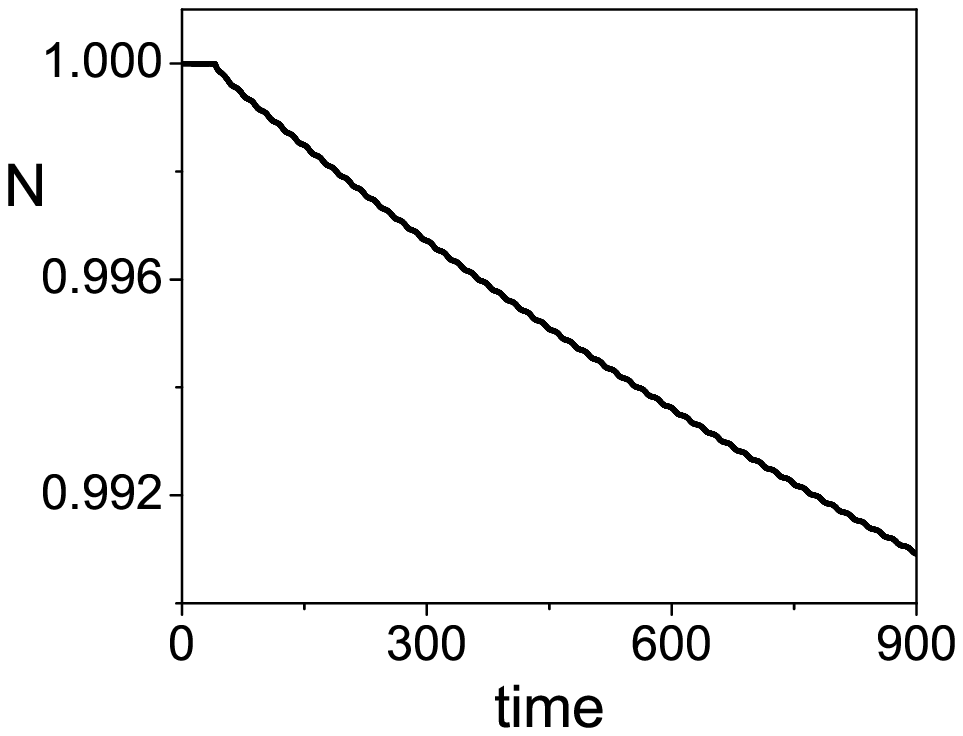} (g)}

 \caption{Configuration of the quasi-stabilized wavefunction. Parameters are the same as in previous figures.
 (a) A snapshot of an amplitude profile. (b) Tiny oscillations in the tail of the quasistabilized solution (c)
Amplitude of the wavefunction far from the origin (the tail plus
outgoing cylindrical wave). (d) Real part of the wavefunction far from
the origin multiplied by $\sqrt{r}$. (e) Snap-shot  of the phase factor
of the quasi-stabilized solution. It can be seen that it is parabolic
only at very small $r$. The curve has an inflection point at $r \le 1$.
f),g) The slowly decaying  norm of the solution. Although the trap was
turned off at $t = T_{off}=30$, the norm
 remains almost constant until the flux of atoms leaking from the core soliton reach the edge of the mesh and begin to
 disappear. After that it decreases slowly.   }
 \label{f1efgh}
\end{figure*}

 In Fig. \ref{f1efgh} configuration of the quasi-stabilized wavefunction is
shown. One can see the smooth core pulse profile, tiny oscillations in
the tail, and an outgoing cylindrical wave leaking from the core pulse.
In Fig. \ref{f1efgh}e the behavior of the phase factor is shown. It is
seen to differ from parabolic with  $r$ considerably.

Figs. \ref{f1efgh}f,g shows the slow decay of the norm of the solution
due to the flux of atoms from the core to infinity.  We made a series
of numerical experiments with different parameters. We found that the
behavior of the matter-wave pulse is often unpredictable. When the
Gaussian approximation predicts stabilization, in the corresponding
numerical solution it does not necessarily occur.  Neither can the
method of moments give reliable predictions for the stabilization. We
checked the latter method carefully. As it was mentioned already in
Sections 1 and 2, the method relies on the crucial approximation of Eq.
(\ref{argu} ). It is due to this approximation one obtains the
existence of dynamical invariants $Q_1$ and $Q_2$ (see Eq.
\ref{quinv}). As a result, dynamics is determined by  Eq.
(\ref{xmodel}). Returning back to Figure \ref{f1efgh}, we see a
snapshot of the phase factor, arg $\psi$, of a stabilized solution. It
clearly demonstrates that none of the PA-based methods reproduce the
dynamics of the system adequately. Only at small $r$ the parabolic law
is fulfilled, while the deviation from this quadratic dependence is
very strong even at $r \le 1$, where the amplitude of the solution is
not small at all (and is sufficient to drastically influence the
dynamics of the system). Snapshots at other  moments produce similar
results: the phase of the solution is changing with time but remains
very far from being parabolic in $r$. It is easy to check that
dynamical properties of the system within a variational approximation
are very sensitive to $r-$ dependence in the phase factor of a trial
function. To check the dynamics further, we calculated time evolution
of the "invariants" $Q_1$ and $Q_2$ in the stabilized solution. They
are constants in the model but not in the exact numerical solution. We
found that in the numerical quasi-stabilized solution these magnitudes
oscillate around some mean value. Actually, it was already found in
Ref. \cite{Garcia} that the method of moments does not work for
Gaussian initial data, still it is interesting to trace dynamics of
relevant magnitudes.  The time evolution of  $Q_1$ and $Q_2$, and other
magnitudes related to the method of moments are shown in Figures
\ref{rt_8ab}, and \ref{rt_8efg}. It is seen that the magnitudes of
$Q_1$ and $Q_2$ related to a stabilized soliton undergo slow
oscillations.

When calculating values of  $Q_1$ and $Q_2$, and other properties of
the quasi-stabilized solution it is necessary to stop integration at
some reasonable value of $r=r_{max}$ (we take $r_{max}=20$ where the
amplitude of the wavefunction becomes very small (of order $10^{-4}$ in
our case). In that way we separate the properties of the
quasi-stabilized soliton from that of the tail which, although has very
small amplitude, can carry large moments $I_2,I_3$ and would give large
contribution to $Q_1$ and $Q_2$ (so that in the corresponding figures
we presented these quantities for the core soliton and the whole
solution (including tail) separately).

Similar features can be seen in Fig. \ref{fg2} where calculations with
$g_0=-7.0$ are presented. Several snapshots of the phase factor at
different moments are presented in order to demonstrate that
non-quadratic behavior of the phase factor is typical. Time evolution
on very long time is traced. We find that sometimes magnitudes of $Q_1$
and $Q_2$ of stabilized solutions are almost conserved (undergoing
small oscillations about its mean value) despite the strongly
non-quadratic behavior of the phase factor. It suggests that the method
of moments developed in \cite{Garcia} might provide useful perspective
for studying the problem and it would be fruitful to extend it taking
into account non-parabolicity of the phase factor.

\section{Concluding remarks}

Despite there are many publications dedicated to the stabilization of a
trapless BEC by the rapidly oscillating scattering length, it seems
that the strong non-parabolic behavior of the phase of the stabilized
wavefunction has not been brought to attention yet. It should be noted
that the role of deviation of  the phase profile of  NLSE solutions
from the parabolic shape was addressed previously in the contexts of
solitons in optical fibers  in Refs.
\cite{nonparabolic1,nonparabolic2}.

Despite that several independent methods were used previously, we have
seen that three of the four theoretical methods used rely on the
unwarranted parabolic approximation, while the fourth method (direct
averaging of GP equation) is, strictly speaking, incorrect, despite its
inspiring motivation (in the sense that the omitted terms has the same
order as those responsible for the creation of the effective potential
). Besides, we find that there is no evidence presently for
stabilization in a strict sense. It seems that the numerical examples
presented so far deal with quasi-stable solutions which slowly decays
due to the leaking of atoms from the core pulse as an outgoing
cylindrical wave. It means that even from a numerical point of view
there are no evidence for true 2D solitons (breathers) yet.

It should be mentioned also that the phenomenon of BEC stabilization
has its counterpart in nonlinear optics. As was studied in Ref.
\cite{Towers}, in the periodically alternating Kerr media the
stabilization of beams is possible. Mathematically, one deals with a
similar NLSE. Instead of the time-dependence of the scattering length
of BEC one has dependence of the media nonlinearity coefficient on the
coordinate z along which a beam propagates:

\be  i u_z + \frac{1}{2} \nabla_{tr}^2 u + \gamma(z) |u|^2 u = 0, \ee

where the diffraction operator $ \nabla_{tr}^2 $ acts on the transverse
coordinate $x$ and $y$. Nonlinearity coefficient $\gamma(z)$ jumps
between constant values $\gamma_{\pm}$ of opposite signs inside the
layers of widths $L_{\pm}$. The analysis of this problem was done using
variational approximation based on a natural Sech ansatz $ U=A(z)
\exp[i b(z) r^2+ i \phi(z)] \mbox{Sech}[r/w(z)]$. However, behavior of
the phase factor was not checked \emph{aposteriori }. We see that it
would be useful to investigate the problem of (2+1)-dimensional
solitons in a layered medium with sign-alternating Kerr nonlinearity in
a greater detail because behavior of the phase factor of the numerical
solution has not been reported yet. Interplay between the phenomenon of
stabilization in Kerr media and BEC was addressed also in Ref.
\cite{Adhikari2} in the context of stabilization of (3+1)- dimensional
optical solitons and BEC in periodic optical-lattice potential (without
addressing the issue of validity of the parabolic approximation).

Returning back to the BEC stabilization, we note that the two main
difficulties should be resolved in the future: the non-trivial behavior
of the argument of the stabilized wavefunction, and the possibility to
stop the leak of atoms from the tail of the solution.

Using several non-Gaussian variational functions, we were able to
determine accurately one of the magnitudes characterizing the
stabilization phenomena: critical nonlinearity $g_{cr}$, but not other
dynamical properties such as the frequency of slow oscillations.

\section{Acknowledgements}

Alexander Itin (A.I.) acknowledges support by the JSPS fellowship
P04315.  A.I. thanks Professor Masahito Ueda and Professor S.V.
Dmitriev for helpful discussions.

 This work was supported in part by Grants-in-Aid for
Scientific Research No. 15540381 and 16-04315 from the Ministry of
Education, Culture, Sports, Science and Technology, Japan.

%\clearpage

%\begin{figure*}
%{\includegraphics[width=6cm]{f1ab}(a)}
% {\includegraphics[width=6cm]{f1ac}(b)}
% \caption{ (a) The oscillations of the monopole moment from previous figure on finer scale.
%  Tiny high frequency oscillations are seen. (b)
% In the Fourier transform, pikes corresponding to driving frequency and slow breathing oscillations are
%seen.} \label{f1bc}
%\end{figure*}

%\begin{figure*}
% \includegraphics[width=6cm]{f1k}
% \caption{Snap-shot  of the phase factor of the quasi-stabilized
% solution. It can be seen that it is parabolic only at very small $r$. The curve has an inflection point at $r \le 1$.}
% \label{f1k}
%\end{figure*}

%\begin{figure*}
%{ \includegraphics[width=6cm]{f1j} (a)}
% {\includegraphics[width=6cm]{f1jj} (b)}
% \caption{ The slowly decaying  norm of the solution. Although the
%trap was turned off at $t = T_{off}=30$, the norm
% remains almost constant until the flux of atoms leaking from the core soliton reach the edge of the mesh and begin to
% disappear. After that it decreases slowly.}
% \label{f1jj}
%\end{figure*}

\begin{figure*}
% \centering
{ \includegraphics[width=6cm]{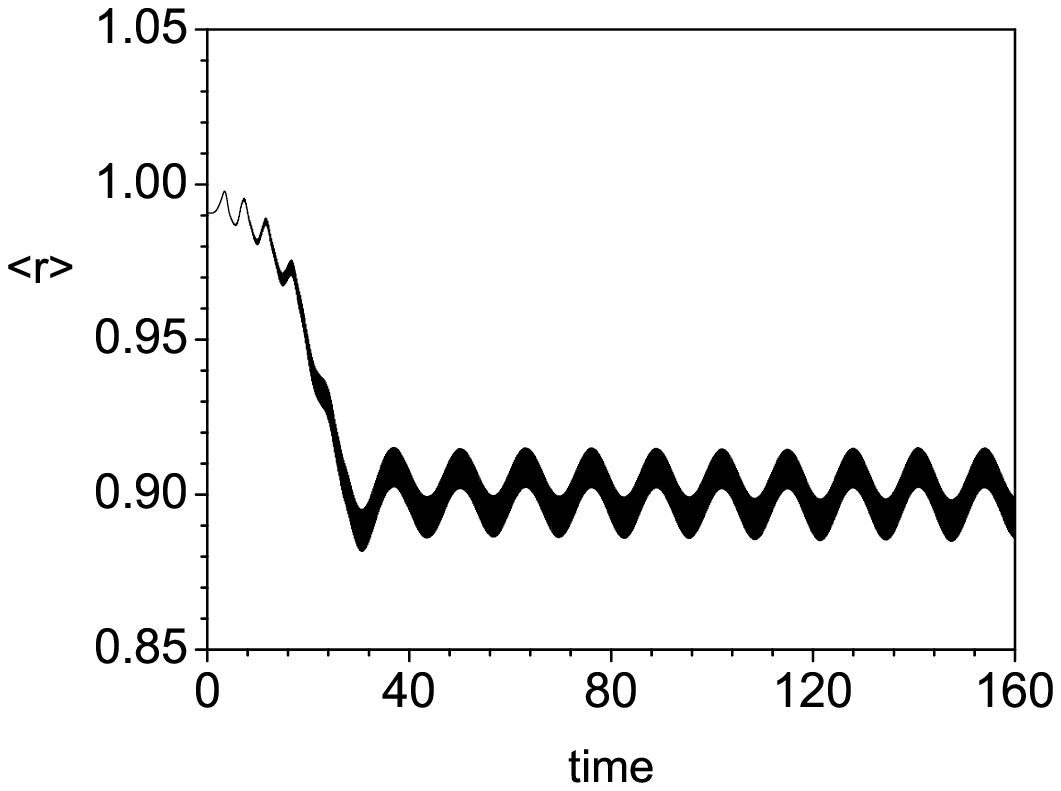} (a)}
{\includegraphics[width=6cm]{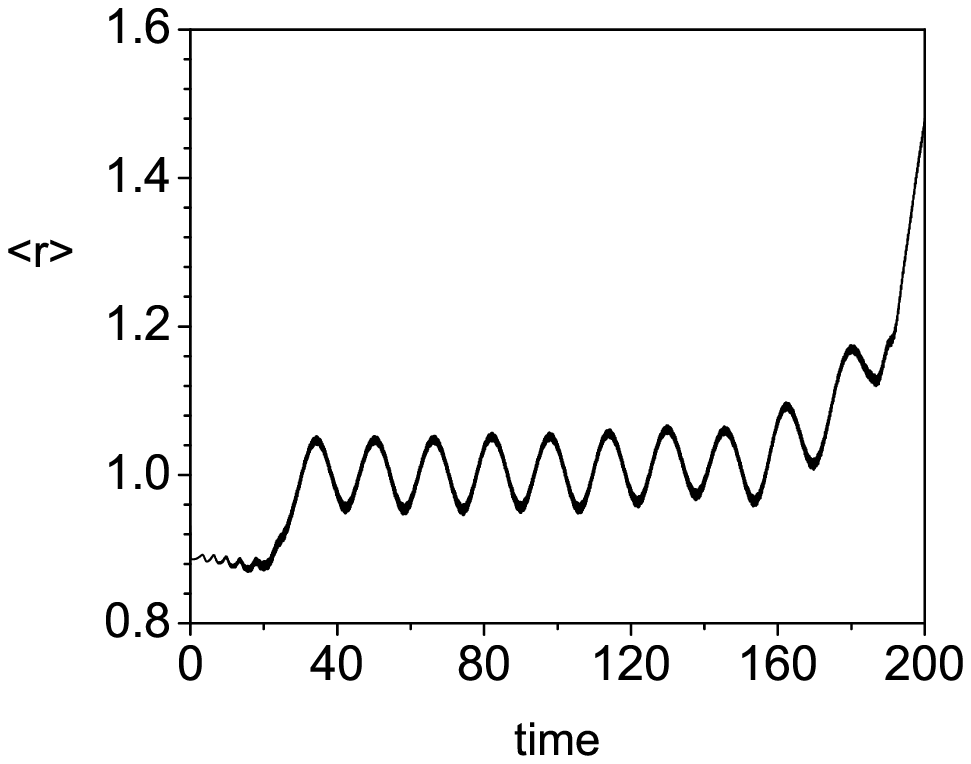} (b) }
 \caption{Oscillations of the monopole moment.
 (a) $g_0=-6.5$, $\Omega=35$, $g_1=10 \pi$ . Initial
 frequency of the parabolic trap is $\omega(0)=0.8$.
 (b) $g_0=-6.5$, $\Omega=30$, $g_1=14.5 $ . Initial
 frequency of the parabolic trap is $\omega(0)=1$. Quasistabilized
 solution is destroyed after several oscillations.}
\label{rt_23}
\end{figure*}

%  rt_2 rt_3 -> rt_23

%\begin{figure*}
% \centering
% \includegraphics[width=6cm]{rt_1c} (a)
% \includegraphics[width=6cm]{rt_1bc} (b)
% \caption{Time evolution of monopole moment. After the trap was  turned off and the nonlinearity turned on,
%   parameter $g_1$ was slowly increased with a linear ramp. After $g_1$ reached a certain threshold value, the
%   solution destabilized rapidly.}
% \label{rt_1}
%\end{figure*}

\begin{figure*}
{\includegraphics[width=6cm]{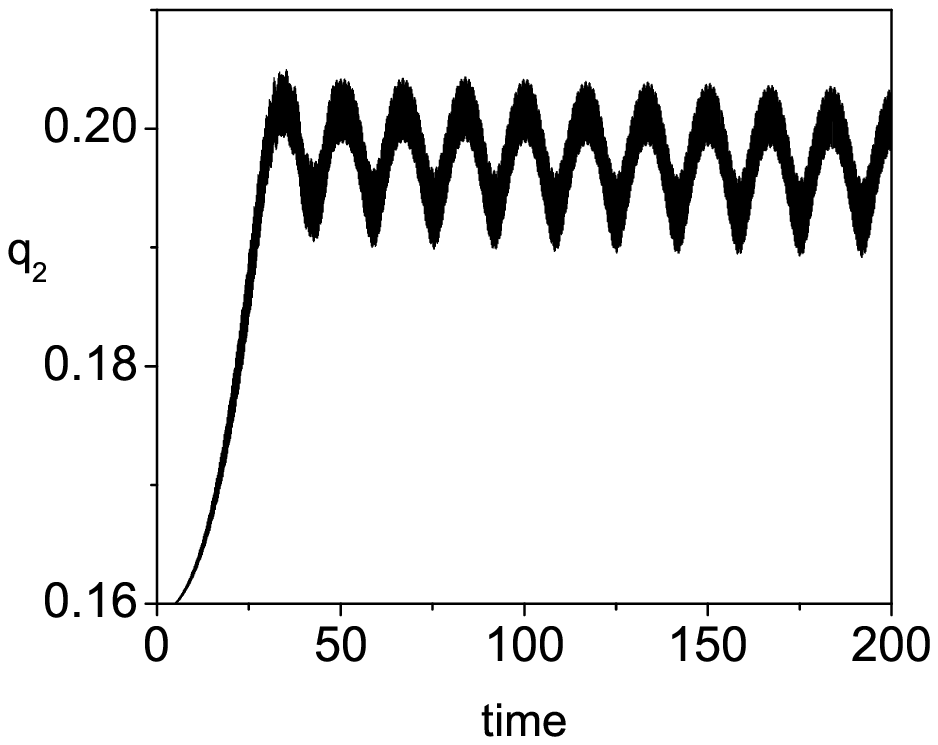} (a)}
{\includegraphics[width=6cm]{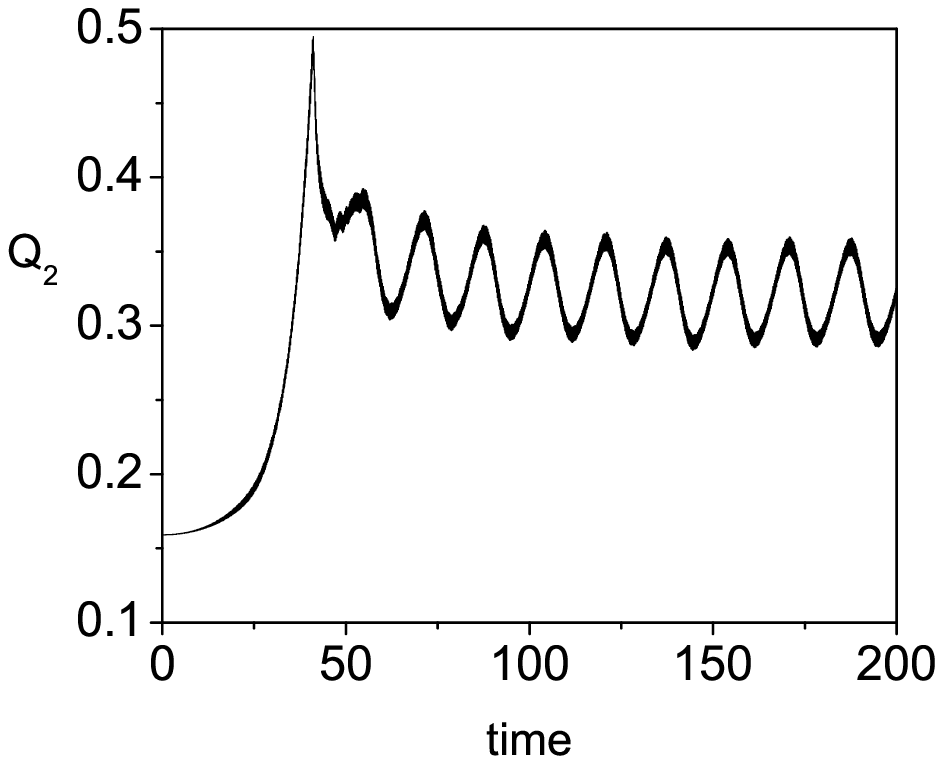} (b)}
{\includegraphics[width=6cm]{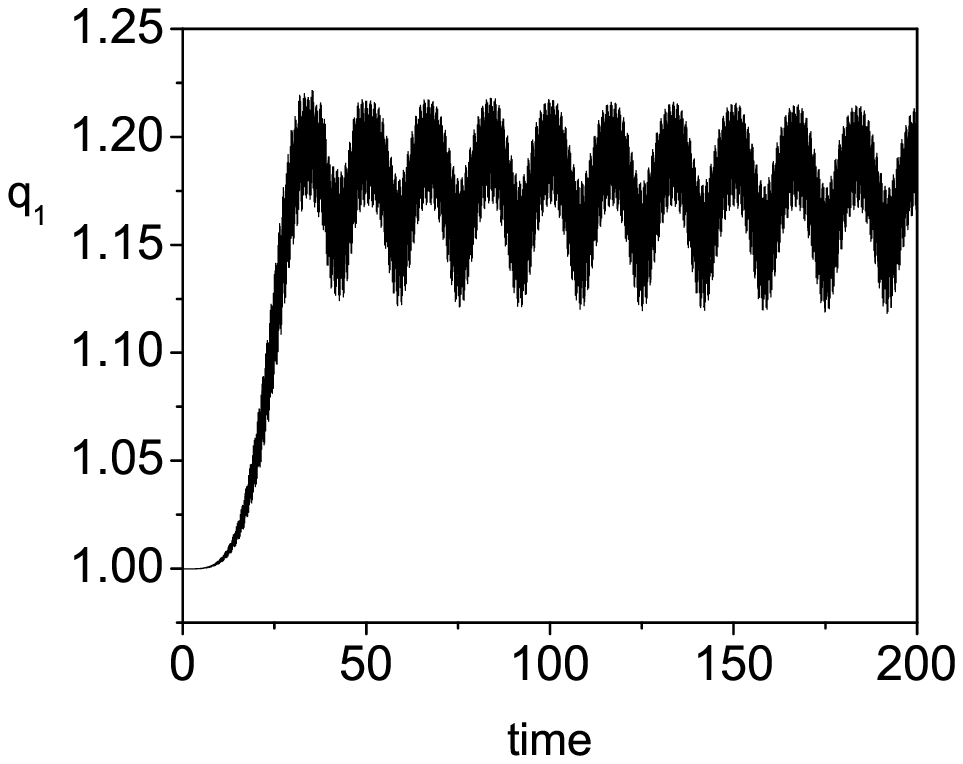} (c)}
{\includegraphics[width=6cm]{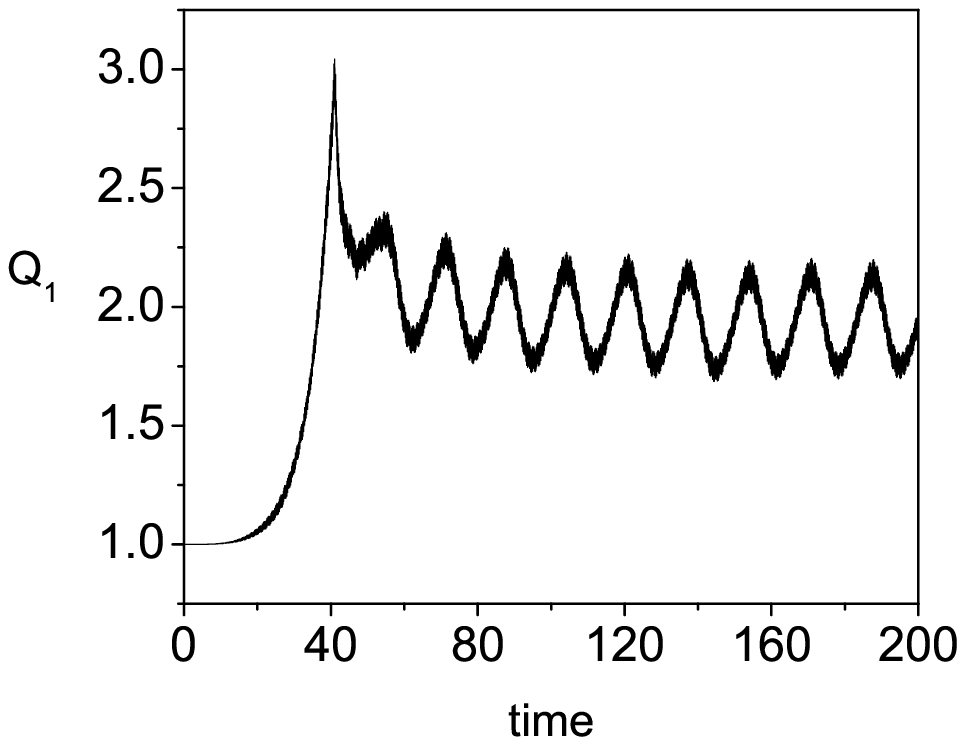} (d)}

%{\includegraphics[width=6cm]{rt_8h} (e)}

\caption{ Time evolution of the integral quantities  $Q_1,Q_2$. (a)
Oscillations of "shortened" $\mbox{Q}_2$ (designated as $q_2$). We
integrate expressions entering Eq. (\ref{quinv}) from $r=0$ to $r
\approx 8$ so that it characterizes the core part of the solution
(quasi-stabilized soliton) without the oscillating tail. (b) Time
evolution of full $Q_2$. The expressions (\ref{quinv}) were integrated
from $r=0$ to $r \approx 120$ so that it includes large contribution
from the oscillating tail. (c) Time evolution of "shortened"
$\mbox{Q}_1$ (designated as $\mbox{q}_1$). We integrate expressions
entering Eq. (\ref{quinv}) from $r=0$ to $r \approx 8$ so that it
characterizes the core part of the solution. Dynamics of the core
soliton for quite a long time is almost independent of the behavior of
the tail which after reaching the edge of the grid begin to disappear.
(d) Time evolution of the full $Q_1$ (including large contribution from
the oscillating tail which depends on location of the absorbing
potential and the mesh size ). }

%  (c) Time evolution of the norm of the solution. When flying away part of the tail
%reaches the edge of the grid with absorbing potential, the norm begin
%to lessen very slowly and steadily. This almost does not influence the
%established oscillations of the monopole moment, see next Figs.}
 \label{rt_8ab}
\end{figure*}

%\begin{figure*}
%{\includegraphics[width=6cm]{rt_8c} (a)}
%{\includegraphics[width=6cm]{rt_8d} (b)} \caption{Oscillations of the
%magnitude of $Q_1$ of the stabilized soliton and the whole solution
%(including weakly interacting tail). (a) Time evolution of "shortened"
%$\mbox{Q}_1$ designated as $\mbox{q}_1$. We integrate expressions
%entering Eq. \ref{quinv} from $r=0$ to $r=20$ so that it characterize
%core part of the solution (stabilized soliton). Dynamics of the core
%soliton for quite a long time is almost independent of the behavior of
%the tail which after reaching the edge of the grid begin to disappear
%(next Figure). (b) Time evolution of full $Q_1$ (including large
%contribution from the oscillating tail which depends on location of the
%absorbing potential and the mesh size ).
%  }
% \label{rt_8cd}
%\end{figure*}

\begin{figure*}
{\includegraphics[width=6cm]{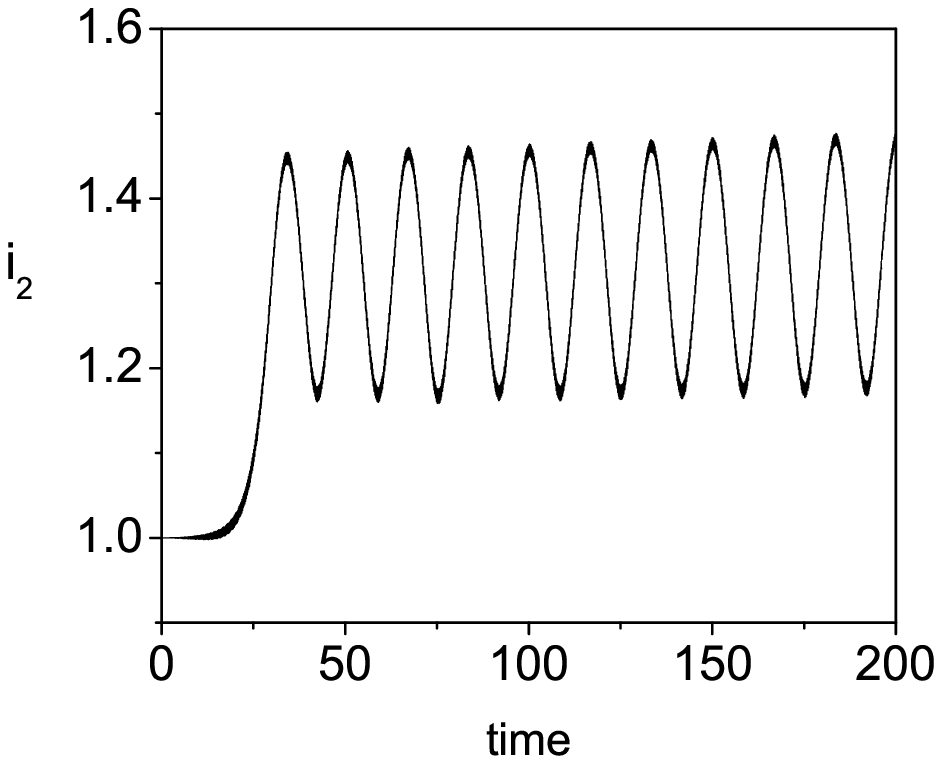}(a)}
{\includegraphics[width=6cm]{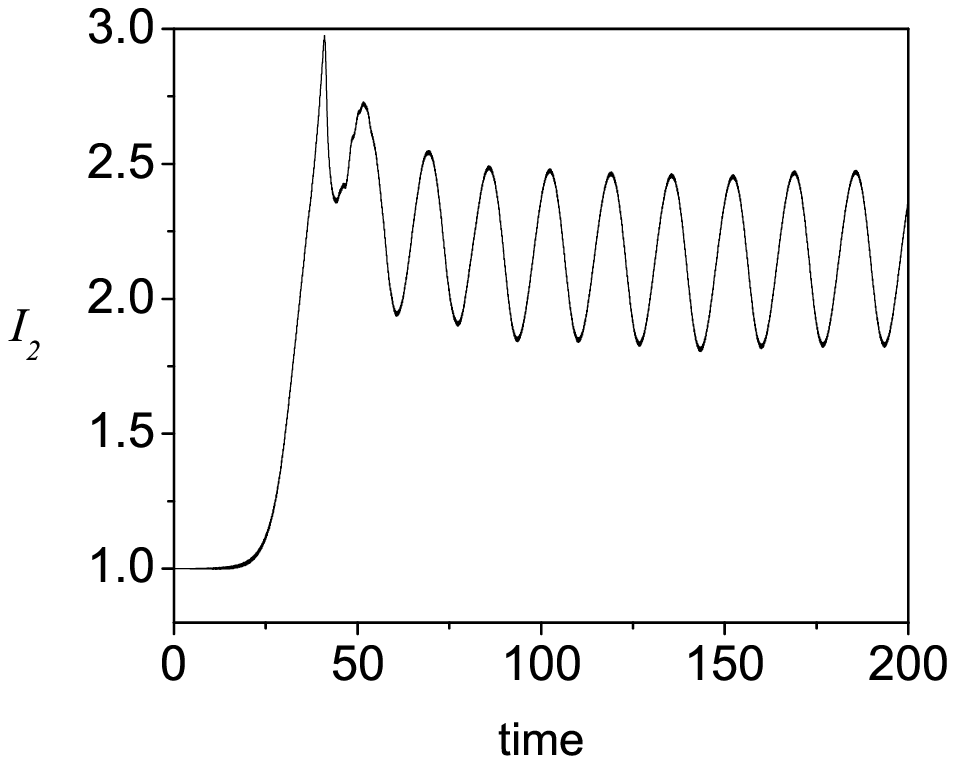}(b)}
{\includegraphics[width=6cm]{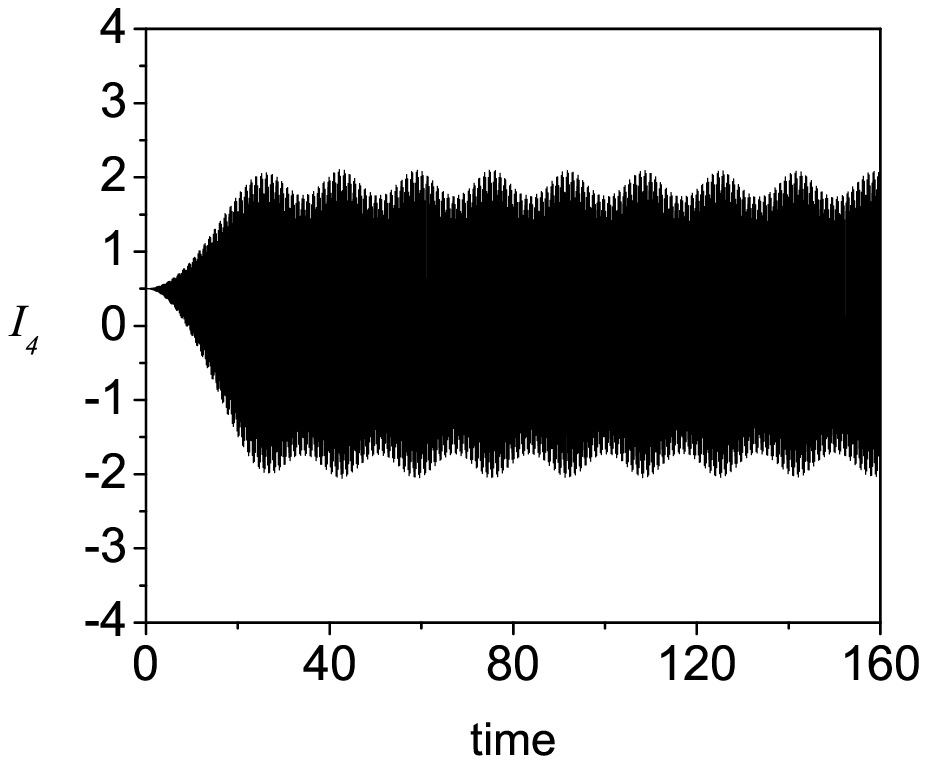}(c)}
{\includegraphics[width=6cm]{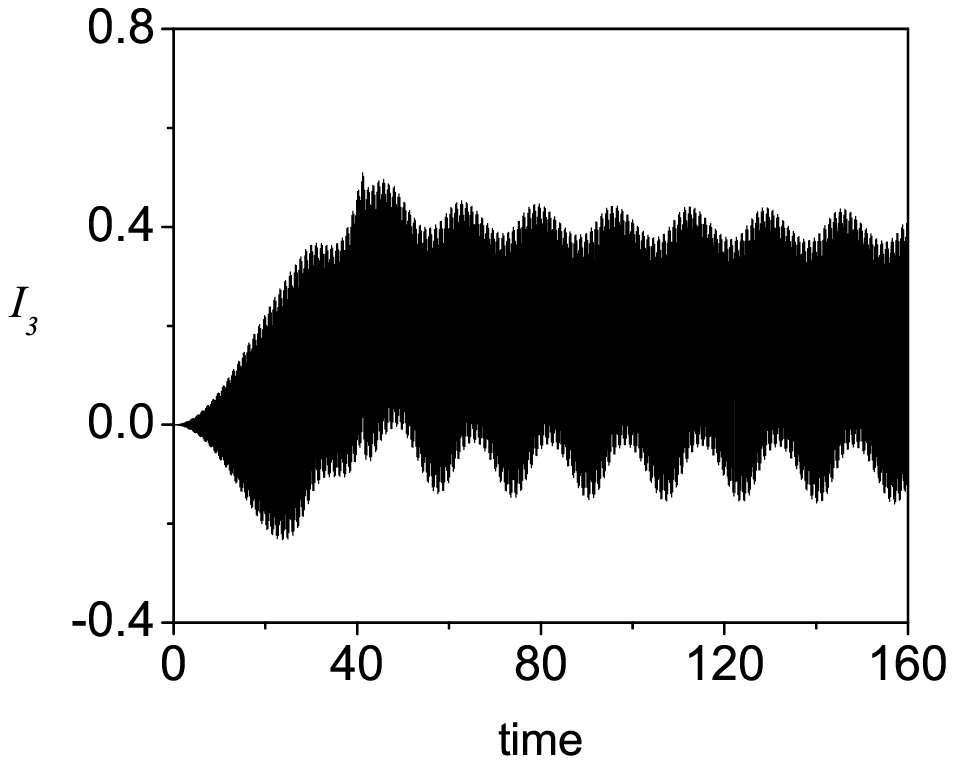}(d)}
 \caption{Time evolution of the
moments ($I_2,I_3,I_4$). (a) Oscillations of the  second moment $<r^2>$
of the core soliton (designated as $i_2$). The boundary of the core of
the quasi-stabilized soliton was taken to be $r \approx 8$.
 (b) Time evolution of the  second moment $I_2=<r^2>$ of the whole solution including tail
  (this magnitude depends on mesh size, here $r_{max} \approx $ 120)
(c) Time evolution of $I_4$. (d) Time evolution of $I_3$. }
\label{rt_8efg}
\end{figure*}

%\begin{figure*}
% \centering
%{\includegraphics[width=6cm]{rt_8k}(a)}
%{\includegraphics[width=6cm]{rt_8j}(b)} \caption{ (a) Time evolution of
%$I_4$. (b) Time evolution of $I_3$}
% \label{rt_8kj}
%\end{figure*}

\begin{figure*}
 \centering
{\includegraphics[width=6cm]{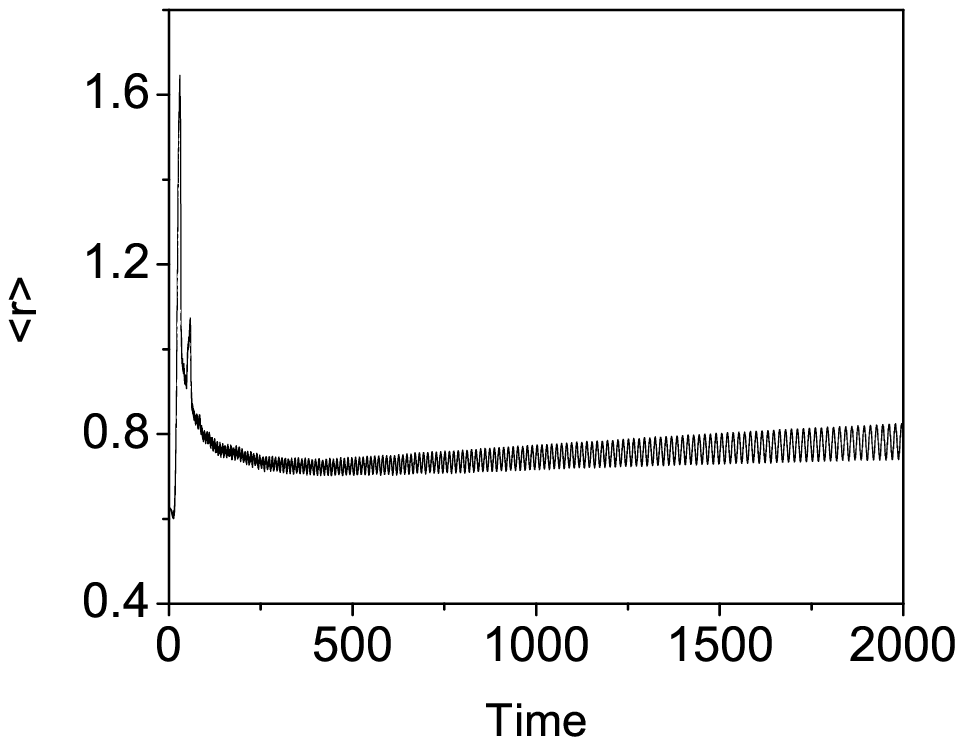} (a)}
{\includegraphics[width=6cm]{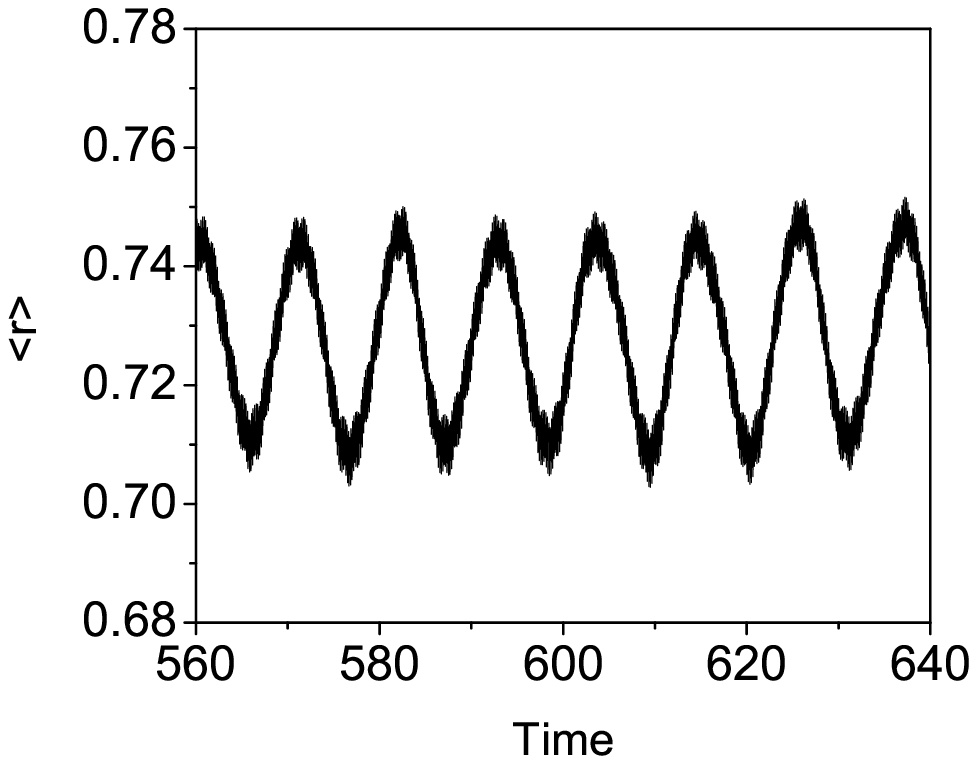} (b)}
{\includegraphics[width=6cm]{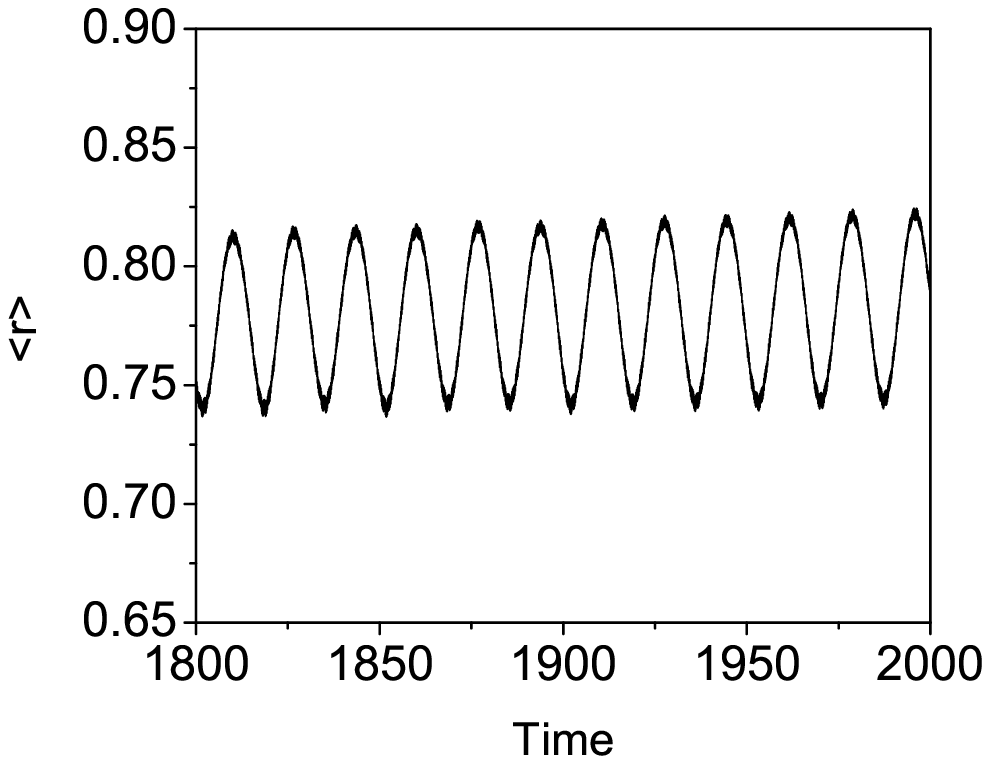}(c)}
{\includegraphics[width=6cm]{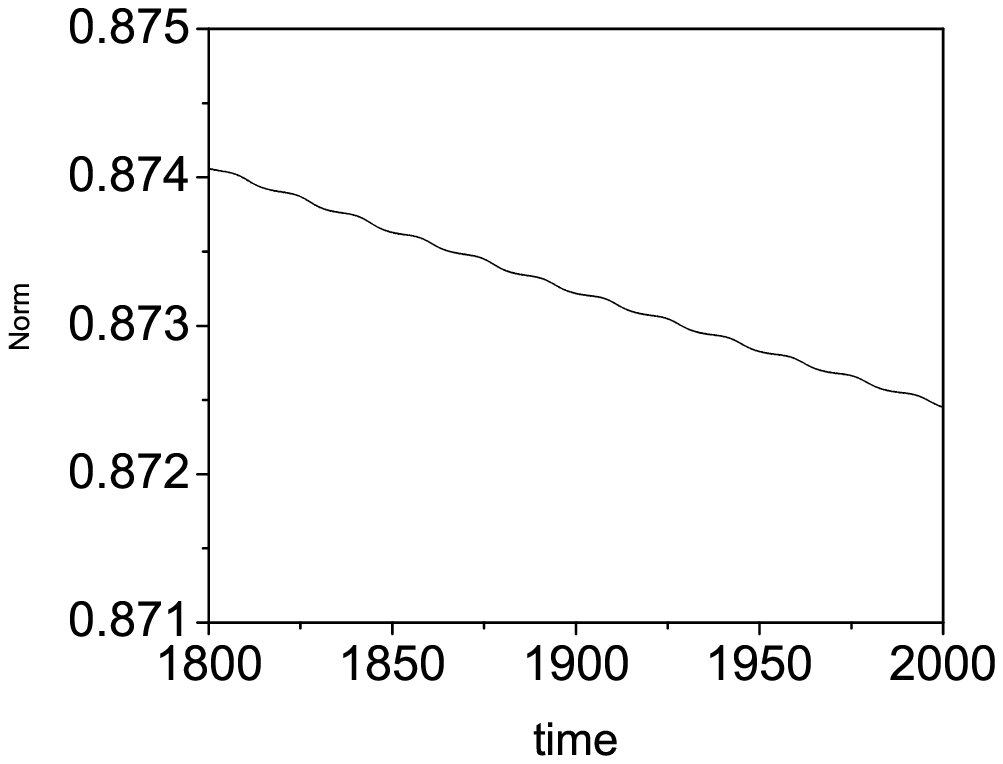}(d)}
{\includegraphics[width=6cm]{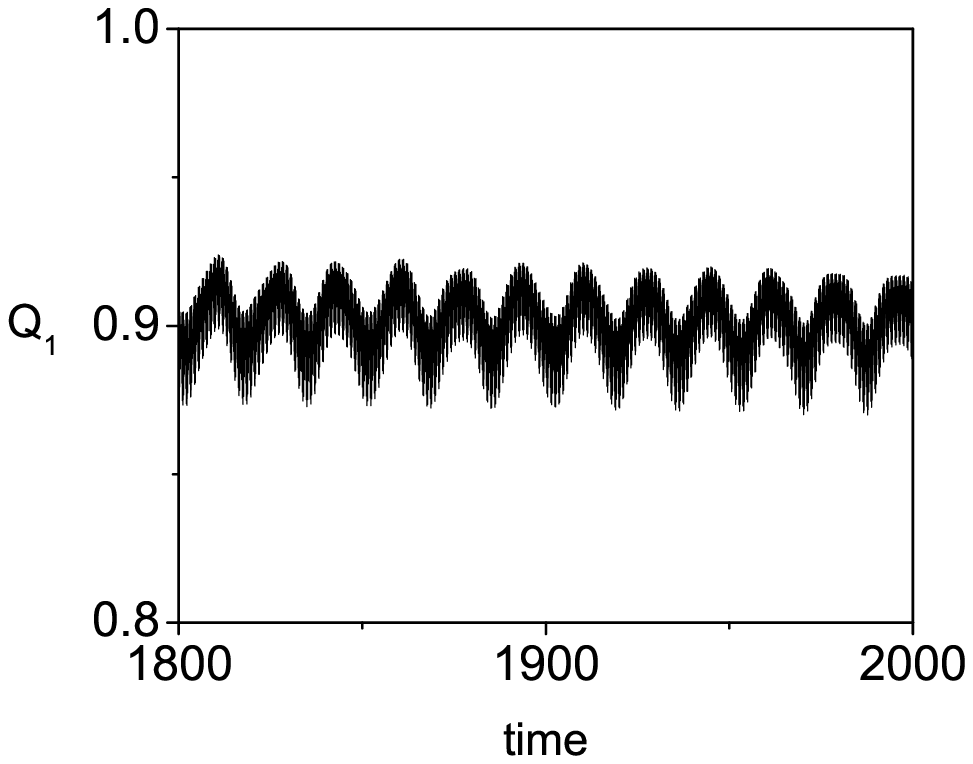}(e)}
{\includegraphics[width=6cm]{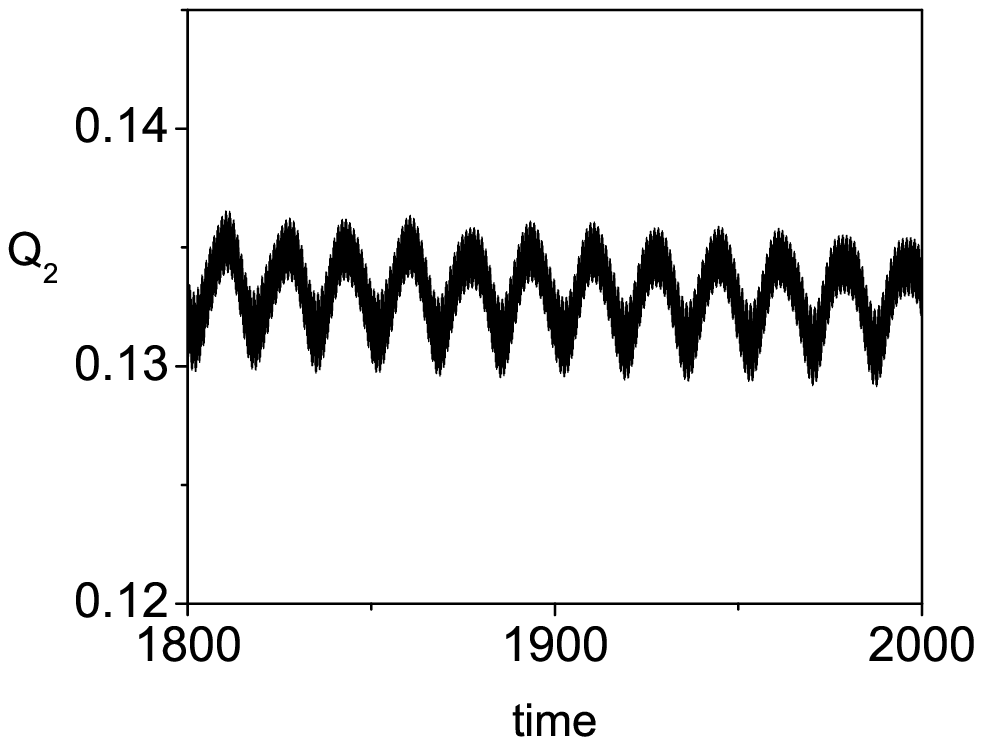}(f)}
{\includegraphics[width=6cm]{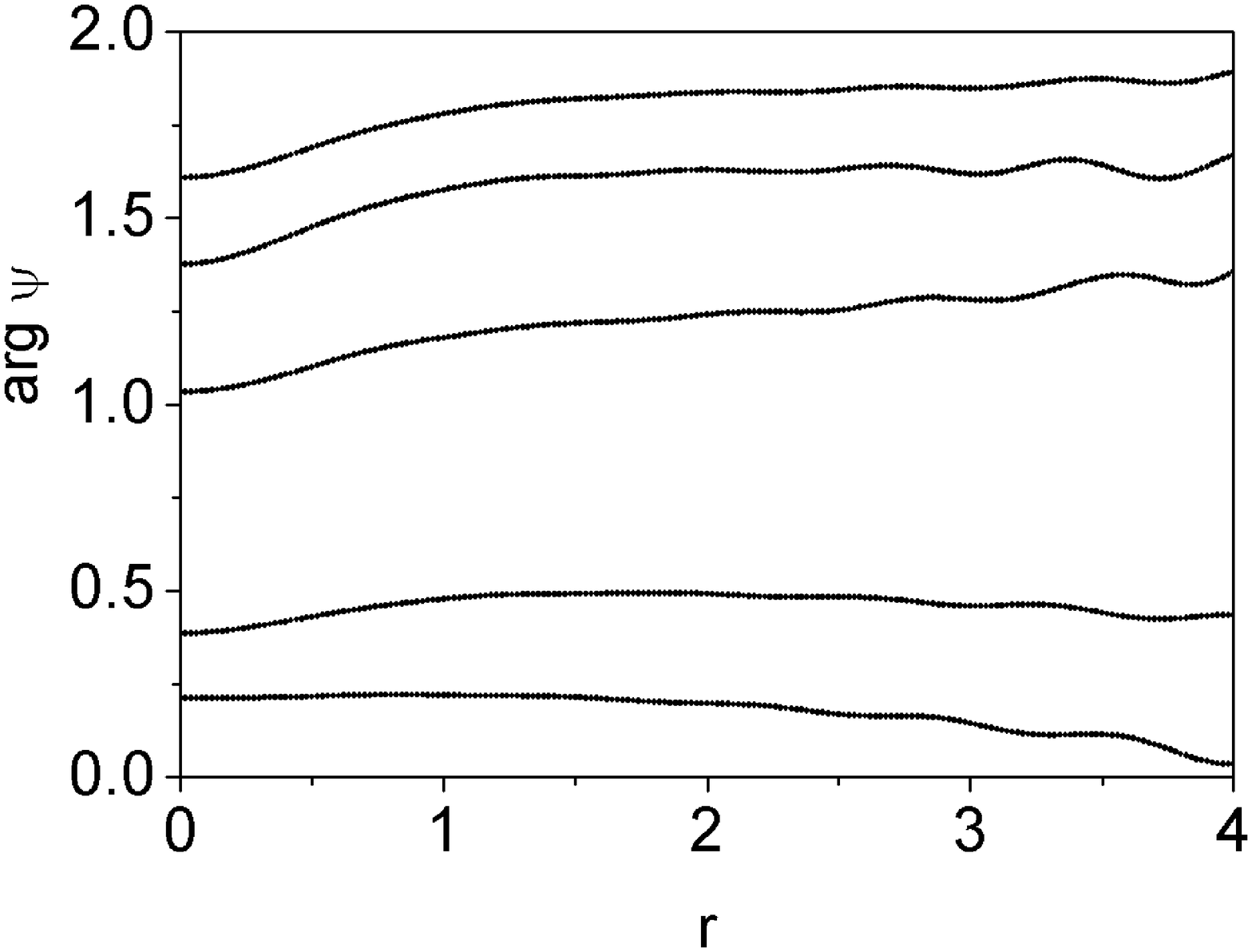}(g)}

\caption{ Oscillations of the monopole moment of a quasistabilized
solution. Parameters are $g_0=-7.0$, $\Omega=40$, $g_1 = 8 \pi$ .
Initial
 frequency of the parabolic trap was chosen to be $\omega(0)=4.0$. (a) Time evolution of the monopole moment  on very long time.
 (b) Detailed picture of the time evolution of the monopole moment of a quasistabilized solution about $t \approx 600$.(c) Detailed picture of the time evolution at $t=1800 \sim 2000$.
 (d) Decaying norm of the solution.
(e) Time evolution of the integral quantity $Q_1$ (calculated for the
core part of the wavefunction) (f) Time evolution of $Q_2$. (g)Several
snap-shots of the phase factor of the quasi-stabilized solution (made
at different moments). Note that typical behavior of the phase factor
is not quadratic with $r$ at all. }
 \label{fg2}
\end{figure*}

%\begin{figure*}
% \centering
%{\includegraphics[width=6cm]{fg2k_c}(a)}
%{\includegraphics[width=6cm]{fg2q10}(b)}
%{\includegraphics[width=6cm]{fg2q20}(c)} \caption{ (a) Several
%snap-shots  of the phase factor of the quasi-stabilized solution (made
%at different moments). Note that typical behavior of the phase factor
%is not quadratic with $r$ at all. (b) Time evolution of $Q_1$
%(calculated for the core part of the wavefunction) (c) Time evolution
%of $Q_2$. }
% \label{fg2k}
%\end{figure*}


\begin{thebibliography}{10}

\bibitem{RefB1}
 F. Dalfovo, S. Giorgini, L. P. Pitaevskii, and S. Stringari, Rev.
Mod. Phys. 71, 463 (1999).


\bibitem{Towers} I. Towers, B. A. Malomed, J. Opt. Soc. Am. B {\bf 19},
537 (2002).


\bibitem{Hau}

L.V. Hau, M. M. Burns, J. A. Golovchenko Phys. Rev. A {\bf 45},
6468–6478 (1992).

\bibitem{SU}
H. Saito and M. Ueda, Phys. Rev. Lett. {\bf 90}, 040403 (2003).

\bibitem{AB}
F. Kh. Abdullaev, J.C. Bronski, and R. M. Galimzyanov,
cond-mat/0205464; Physica D {\bf 184},319 (2003).


\bibitem{Abdullaev1}
F. Kh. Abdullaev, J.G. Caputo, R. A. Kraenkel, B.A. Malomed, Phys.Rev.
A {\bf 67}, 013605 (2003).

\bibitem{Abdullaev2}
F. Kh. Abdullaev, A.Gammal, L.Tomio, T. Frederico, Phys. Rev. A {\bf
63}, 043604.

\bibitem{Garcia} G. D. Montesinos, V.M. Perez-Garcia, P.J. Torres,  Physica {\bf D} 191, 193–210 (2004).
\bibitem{vektor} Gaspar D. Montesinos et. al, Chaos 15, 033501 (2005).

\bibitem{Adhikari}  S.K. Adhikari, Phys. Rev. A {\bf 69}, 063613 (2004).


\bibitem{3D} M. Matuszewski et. al, Phys. Rev. Lett. 95, 050403 (2005).

\bibitem{Metens}
C. Huepe, S. Metens, G.Dewel, P. Borckmans, M.E. Brachet, Phys. Rev.
Lett. {\bf 82}, 1616.




\bibitem{Kagan} Y. Kagan, E.L. Surkov  and G.V. Shlyapnikov, Phys. Rev. A 54,
R1753 (1996)


\bibitem{Kath} T.S. Yang and W. L. Kath, Optica Letters {\bf 22},
985(1997).

\bibitem{Ref_Garcia} J. Lei, M. Zhang, Lett. Math. Phys. {\bf 60}, 9
(2002).

\bibitem{Cirac}
V.M. Pe´rez-Garcia, H. Michinel, J. I. Cirac, M. Lewenstein, and P.
Zoller, Phys. Rev. Lett. {\bf 77}, 5320 (1996); Phys. Rev. A {\bf 56},
1424 (1997).


\bibitem{nonparabolic1}

Opt.Commun. {\bf 147}, 317 (1998).

\bibitem{nonparabolic2}

Progr. Optics {\bf 43}, 71 (2002).



\bibitem{Pramana} D. Anderson, M. Lisak, A. Bertson, Pramana – J. Phys. {\bf 57}, 917 (2001).


\bibitem{AKN}
V.I.Arnold, V.V.Kozlov, and A.I.Neishtadt, {\it Mathematical aspects of
classical and celestial mechanics } { (second edition, Encyclopaedia of
mathematical sciences 3, Springer-Verlag, Berlin, 1993)}.
\bibitem{It1}
See, for example: A.P.Itin, A.I.Neishtadt, A.A.Vasiliev, Phys. Lett. A.
291, 133 (2001); A.P. Itin, R. de la Llave, A. I. Neishtadt, A. A.
Vasiliev, Chaos 12, 1043 (2002); A.P. Itin, A.A. Vasiliev, A.I.
Neishtadt, Physica D 141, 281 (2000).

\bibitem{Adhikari2} S. K. Adhikari, Phys. Rev. E 71, 016611 (2005).
\end{thebibliography}
\end{document}